\documentclass[reprint,aps,prb,floatfix]{revtex4-2}

\usepackage{amsmath,amssymb,bm,stmaryrd}
\usepackage{graphicx,epstopdf}
\usepackage{array,multirow,dcolumn}
\usepackage{color,xcolor}
\usepackage{subfigure}
\usepackage{float}
\usepackage{natbib}
\usepackage[T1]{fontenc}
\usepackage[utf8]{inputenc}
\usepackage[english]{babel}
\usepackage{hyperref}
\usepackage[normalem]{ulem}
\usepackage{soul}
\usepackage{textcomp}

\hypersetup{
	colorlinks=true,
	linkcolor=purple,
	citecolor=blue,
	urlcolor=purple
}

\begin{abstract}
Symmetry offers a useful approach to unfold the intertwined degrees of freedom. Thus it paves the way to resolve
coexisting quantum orders into distinct symmetry sectors. Motivated by the recent observation of superconductivity in nano-flaked IrTe$_2$, we investigate the superconductivity in strain-stabilized 
two-dimensional (2D) limit of IrTe$_2$ by combining density-functional theory with mean-field solution of spin-fluctuation mediated pairing interaction on a symmetry-constrained ${\bf k}\cdot{\bf p}$ model. The spin-orbit coupled 
band structure shows $\Gamma$-centred Fermi sheets with coexistence of band-selective Rashba-like (in-plane) and Ising-like (out-of-plane) superconductivity. Remarkably, the superconducting gaps are odd in spin, orbital, and momentum channels despite the presence of global inversion symmetry. Fermi surface topologies and little-group symmetry enforce distinct irreducible representations to the Rashba and Ising channels, forbidding their mixing. Our findings open up a symmetry-based route to multichannel superconductivity in 2D transition-metal dichalcogenides with unique functionalities.
\end{abstract}
\begin{document}
%\title{Inversion-odd Singlet Type-II Ising Superconductivity  in the Centrosymmetric IrTe$_{2}$}
\title{Coexistence of Rashba and Ising Spin-Singlet Pairings in Two-Dimensional IrTe$_{2}$}

\author{Kunal Dutta$^{1}$} 
\author{Rajesh O. Sharma$^{2,3}$}
\author{Shreya Das$^{4,5}$}
\author{Indra Dasgupta$^{1}$}  \email{sspid@iacs.res.in}   
\author{Tanmoy Das$^{2}$} \email{tnmydas@iisc.ac.in}
\author{Tanusri Saha-Dasgupta$^{4}$} \email{t.sahadasgupta@gmail.com}
\affiliation{$^{1}$School of Physical Sciences, Indian Association for the Cultivation of Science, 2A and 2B Raja S.C. Mullick Road, Jadavpur, Kolkata 700032, India.\\
$^{2}$ Department of Physics, Indian Institute of Science, Bangalore 560012, India.\\
$^{3}$ Department of Science and Humanities (Physics), Dr. B. R. Ambedkar Institute of Technology, Dollygunj, Sri Vijaya Puram 744103, Andaman and Nicobar Islands, India.\\
$^{4}$ Department of Condensed Matter and Materials Physics, S. N. Bose National Centre for Basic Sciences, JD Block, Sector III, Salt Lake, Kolkata, West Bengal 700106, India.\\
$^{5}$ Masaryk University, Kamenice 753/5, 625 00 Brno, Czech Republic.
}
\date{\today}
\pacs{}
\maketitle

%%%%%%%%%%%%%%%%%%%%%%%%%%%%%%%%%%%%%%%%%%%%%%%%%%%%%%%%%%%%%%%%%%%%%%%%%%%%%%%%%%%%%%%%%%%%%%%%%%%%%%%%%%%%
%\section{\label{sec:level1}Introduction}
\noindent
\textit{Introduction-} Recently, the extraordinary properties driven by the interplay between emergent spin-orbit coupling (SOC) and crystalline symmetry in 2D materials have attracted considerable attention~\cite{PhysRevLett.108.196802,Advanced_Materials_2021,Huang2017,Gong2017,RevModPhys.83.1057}. A prominent example is the recent discovery of a novel Ising superconductivity in 2D noncentrosymmetric materials 
like 2H phase of transition metal dichalcogenides (TMDs) ~\cite{Saito2016,delaBarrera2018,PhysRevX.10.041003} where  upper critical field ($B_{c2}$) surpasses the Pauli limit ~\cite{LI2021100504,Zhang_2021}. In this Type-I Ising superconductivity, strong SOC induces Zeeman-like spin splitting that reverses between the $K$ and $K'$ valleys, leading to spin–valley locking and a pronounced out-of-plane spin polarization of Cooper pairs ~\cite{doi:10.1126/science.aab2277,Xi2016,doi:10.1073/pnas.1716781115,LI2021100504,Zhang_2021,PhysRevX.8.021002}. Lack of inversion symmetry, though, mixes singlet and triplet components, posing challenge for spin-selective transport \cite{SigristRMP}. Very recently, an alternative form, known as Type-II Ising superconductivity has been reported in centrosymmetric 2D materials with $D_{3d}$ point-group symmetry~\cite{doi:10.1126/science.aax3873,PhysRevLett.123.126402}. Here, multifold band degeneracies protected by crystal and time-reversal symmetries produce spin–orbital locking, rather than spin–valley locking. This results in opposite out-of-plane spin splitting for the degenerate partners.
Based on the guiding principle of SOC-induced band splittings around high-symmetry points,
a number of Type-II Ising superconductors has been predicted ~\cite{doi:10.1126/science.aax3873,PhysRevLett.123.126402,PhysRevB.109.165428,Liu2020} in the parameter space of strength of SOC versus carrier density, though SC properties or the predicted density has not been verified. 

In this letter, we focus on the unexplored case of a monolayer of IrTe$_2$. Bulk IrTe$_2$ with 
van der Waals stacked layers of edge-shared IrTe$_6$ octahedra, is characterized by a poorly understood charge-density-wave (CDW) 
transition at $\sim$260 K that is accompanied by strong diamagnetism and structural anomaly \cite{PhysRevLett.108.116402}. The CDW and SC phases compete with each other, with superconductivity appearing upon suppression of the former  through the Pd or Cu intercalation or Pd doping \cite{PhysRevLett.108.116402,Phase_IrTe2,PhysRevB.87.180501,PhysRevB.97.205142}. On another route, employing the principle of slow ordering kinetics in reduced dimensions, CDW was avoided in IrTe2 nano-flakes, stabilizing SC \cite{Yoshida2018}. This fuels the possibility of the realisation of Ising SC in 2D IrTe$_2$, given the fact that both Ir and Te being 5$d$ and 5$p$ elements, respectively host strong SOC. Considering the mono-layer of IrTe$_2$, we discover that this centrosymmetric 2D material exhibits a novel form of superconductivity that is unusual on several counts: (a) The SC gaps on all Fermi sheets are odd under the exchange of spin, orbital, and momentum which track the underlying spin textures induced by Rashba and Ising SOCs. (b) These Rashba and Ising pairings coexist but do not mix, allowing for the clean disentanglement of in-plane versus out-of-plane spin pairings into distinct bands. The separation of channels unlocks technology-facing advantages: spin-filtered transport, spin-sensitive Josephson interferometry, and strongly anisotropic upper critical fields $B_{c2}$ with robust out-of-plane enhancement in the Ising pairing channel.
This further opens up a broader landscape of possible realization of intriguing symmetry-driven non-trivial superconductivity in the general class of 2D TMDs in the parameter space of filling, SOC, and crystal symmetry.

%%%%%%%%%%%%%%%%%%%%%%%%%%%%%%%%%%%%%%%%%%%%%%%%%%%%%%%%%%%%%%%%%%%%%%%%%%%%%%%%%%%
\begin{figure}[t]
\centering
    \includegraphics[height= 69 mm, width=85 mm,keepaspectratio]{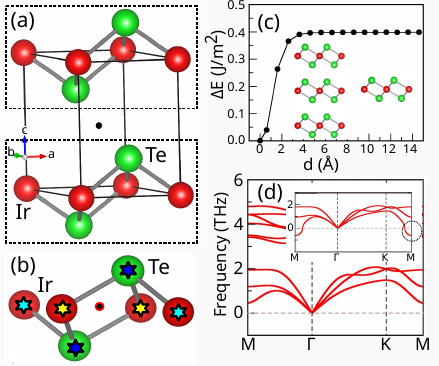}
     \caption{ (a) Crystal structure of bulk IrTe$_2$. The black circle
     represents the inversion center of the structure. The monolayer IrTe$_2$ blocks (marked as dotted boxes) are stacked along the vertical $c$-axis. (b) Crystal structure of monolayer IrTe$_2$. Differently colored stars indicate inversion-symmetric atomic partners, while the black circle represents the inversion center. (c) Calculated energy cost in separating the monolayer blocks as function of interlayer distance, $d$. The converged value of the energy cost at large $d$ provides the estimate of cleavage energy. The inset pictorially illustrates the transition from bulk to monolayer. (d) Phonon spectrum of the monolayer under 1$\%$ tensile biaxial strain. The unstrained monolayer is found to be dynamically unstable, as indicated by the presence of unstable phonon frequencies in the spectrum shown in the inset.}
    \label{Fig:1}
\end{figure}
%%%%%%%%%%%%%%%%%%%%%%%%%%%%%%%%%%%%%%%%%%%%%%%%%%%%%%%%%%%%%%%%%%%%%%%%%%%%

%%%%%%%%%%%%%%%%%%%%%%%%%%%%%%%%%%%%%%%%%%%%%%%%%%%%%%%%%%%%%%%%%%%%%%%%%%%%%%%%%%%
\begin{figure}
	\includegraphics[height= 85 mm, width=75 mm,keepaspectratio]{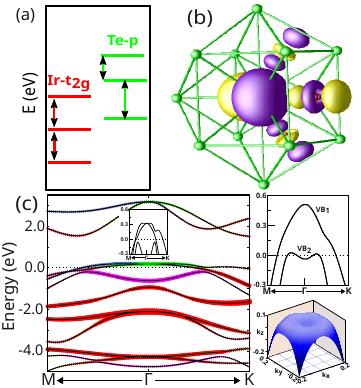}
	\caption{(a) Energy level positions of Te-$p$ and Ir-$t_{2g}$ in IrTe$_2$.
    (b) Te-$p_x$ Wannier function centered at Te site (marked as green ball) within a cage of Ir (marked as red ball) and Te. Plotted is constant value surface with yellow and magenta denoting two opposite signs of the function. (c) Left: Orbital-projected GGA band structure of the IrTe$_2$ monolayer along the high-symmetry path M–$\Gamma$–K–M. The Fermi level is set to zero. Colors denote orbital characters: Te-$p_x$ (green), Te-$p_y$ (blue), Te-$p_z$ (magenta), and Ir-$d$ (red). Zoomed-in views of the band structure near the Fermi level, highlighting
    degenerate Te $p_x/p_y$ bands at $\Gamma$ point is shown as inset.  Right, top: Zoomed-in views of the band structure near the Fermi level upon inclusion of SOC, which splits
    Te $p_x/p_y$ bands at $\Gamma$ point. Right, bottom: GGA+SOC low energy band structure
    in $k_x$-$k_y$ plane.}
	\label{Fig:2}
\end{figure}
%%%%%%%%%%%%%%%%%%%%%%%%%%%%%%%%%%%%%%%%%%%%%%%%%%%%%%%%%%%%%%%%%%%%%%%%%%%%

%%%%%%%%%%%%%%%%%%%%%%%%%%%%%%%%%%%%%%%%%%%%%%%%%%%%%%%%%%%%%%%%%%%%%%%%%%%%%%%%%%%%%%%%%%%%%%%%%%%%%%%%%%%%%
\noindent
\textit{Structural Details-} The non-CDW phase of bulk IrTe$_2$ crystallizes in centrosymmetric, trigonal space group of $P\bar{3}m1$ with Ir layer sandwiched between two triangular layers of Te atoms, forming a monolayer of IrTe$_2$. The 2D monolayers of IrTe$_2$ are stacked on top of each other (cf Fig.~\ref{Fig:1}(a)). Given the van der Waals nature of stacking, the block of IrTe$_2$ monolayer, as illustrated in Fig.~\ref{Fig:1}(b), can be cleaved by mechanical exfoliation. For this an energy barrier needs to be overcome, known as the cleavage energy, defined as $\Delta$E = 2$\frac{(E_{slab}  -  E_{bulk})}{2A}$  where $E_{slab}$ is the total energy of the cleaved system with two exposed surfaces (while the individual layers are sufficiently far to be out of the range of dispersive interactions) and $E_{bulk}$ is the
total energy of the same in the bulk, $A$ being the surface area. To estimate 
$\Delta$E, the surface area-normalized total energy difference between the layer-separated and bulk was calculated by gradually increasing separation  between 
monolayer IrTe$_2$ blocks in the range 0.2–10 Å (cf. Fig.~\ref{Fig:1}(c)), employing the plane-wave basis set  within the framework of density functional theory (DFT) as implemented in Vienna ab-initio Simulation Package (VASP) \cite{PhysRevB.47.558,PhysRevB.54.11169,PhysRevB.50.17953,PhysRevLett.77.3865}. See S1 of supplementary materials (SM) \cite{supplemental} for calculation details. 
    The converged energy difference, provides the measure of cleavage energy, estimated to be 0.39 J/m$^2$, comparable to that of graphene (0.37 J/m$^2$) \cite{Wang2015}.
The 2D structure derived out of the bulk structure, though, exhibit instability of
the quadratically dispersing acoustic ZA mode of the phonon dispersion at M point, as shown in inset of Fig.~\ref{Fig:1}(d). This very soft mode,
which propagates parallel to the layer, corresponds to the layer-bending or ripple mode, known as flexural phonon mode, a characteristic of 2D geometry.
In literature, bi-axial strain has been discussed \cite{flexural} as plausible route to influence  the frequencies of the flexural phonon modes, with tensile strain making it more rigid and compressive strain making it further softer. Following this, we find that the application of 1$\%$ tensile strain makes the 2D structure dynamically stable, as shown in Fig.~\ref{Fig:1}(d). The structure
is also found to be thermodynamically stable at room temperature (see S2 of SM \cite{supplemental}). The monolayer of IrTe$_2$ retains centrosymmetry, since the $D_{3d}$  point group around Ir contains inversion operator, with
inversion-symmetric atomic pairs (cf. Fig.~\ref{Fig:1}(b)). Reduction of dimensionality from bulk to monolayer, though shifts the inversion
center from between the monolayer blocks to that at the mid-point of the Ir layer.

\begin{figure}[t]
	\centering
	\includegraphics[height= 108 mm, width=82 mm,keepaspectratio]{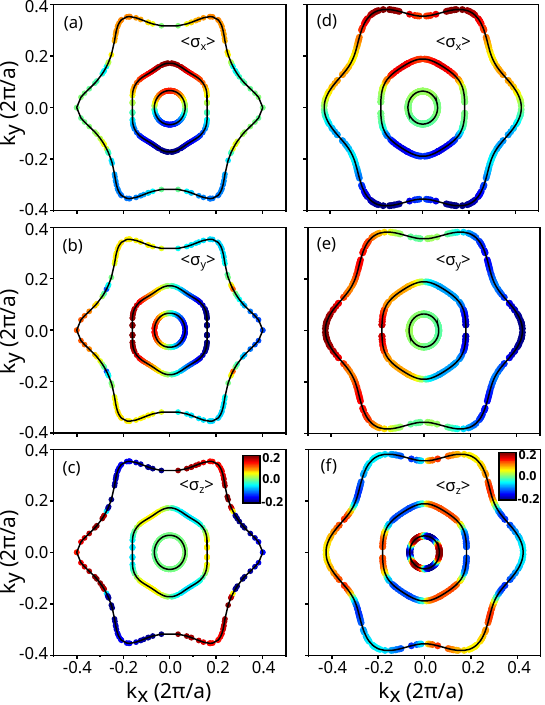}
	\caption{ 
		(a)–(c) Projection of spin expectation values on the Fermi surface, obtained from the DFT calculations. 
        (d)-(f) The same, but obtained from the $k \cdot p$ model Hamiltonian. 
		The strength of the spin expectation values is indicated by the color band shown in (c) and (f). 
	}
	\label{Fig:3}
\end{figure}

\noindent
\textit{Electronic Structure -} The nominal valence of Ir in
 non-CDW phase is suggested to be 3+ with filled shell $t_{2g}^6$ configuration \cite{valence1, valence2}. This implies an average valence of 1.5- of Te, forming
 ligand holes. This scenario is supported by the energy level diagram
 of Ir-$t_{2g}$ and Te-$p$, shown in Fig.~\ref{Fig:2}(a), obtained from the onsite matrix elements of the real space Hamiltonian in Ir-$t_{2g}$ and Te-$p$ Wannier basis (see S3 of SM \cite{supplemental}). Te-$p$ states lie higher than Ir-$t_{2g}$ states, resulting in a negative charge transfer energy scenario in IrTe$_2$. Further, the strong Ir-Te covalency due to extended nature of Ir 5d and Te 5p orbitals is evident from the 
 plot of the Wannier function in Te-$p$ only basis. Representative 
 Te-$p_x$ Wannier function, shown in Fig.~\ref{Fig:2}(b),  
 acquires substantial Ir-d character in its tail. The low-energy band
 structure of monolayer is thus dominated by mixed Ir-$d$-Te-$p$ 
 orbital contribution, as shown in left panel of Fig.~\ref{Fig:2}(c)). 
 In particular, the top three valence bands are composed of anti-bonding Te-$p$ states,
 with dominant Te-$p$ character at $\Gamma$ point, following the negative charge transfer scenario. This is followed by three filled bands of
 significant Ir-$d$ composition, in agreement with nominal Ir$^{3+}$ valence, and subsequently the bonding partner  of Te-$p$ states (see S4 of SM \cite{supplemental}). 
 Interestingly, degenerate Te $p_x/p_y$ appear in the top two valence bands (see inset), protected by $C_3$ and time reversal symmetry of the $D_{3d}$ space group\cite{PhysRevLett.123.126402}. This is followed by a low-lying
 $p_z$ state. In presence of SOC, the four-fold degeneracy of $p_x/p_y$ at top of valence band, gets lifted into two-fold generate 
 $J_z$ = $\pm$ 3/2 (VB1) and $\pm$ 1/2 (VB2) of $J$ = 3/2 manifold (cf Fig.~\ref{Fig:2}(c) right panel). The resultant Fermi surface (FS) comprises of twofold degenerate three sheets, as detailed in discussion pertinent to Fig.~\ref{Fig:3}, with two inner sheets of small volume, and an outer sheet of large volume.
 Among these sheets, the outer one exhibits a pronounced hexagonal, snowflake-like Fermi surface. The two inner sheets are more circular, originating from the inner bands (VB2), exhibiting 
 characteristic Mexican-hat (or camel-back) type dispersion around the Fermi level (cf. right, bottom of Fig.~\ref{Fig:2}(c)).

%%%%%%%%%%%%%%%%%%%%%%%%%%%%%%%%%%%%%%%%%%%%%%%%%%%%%%%%%%%%%%%%%%%%%%%%%%%%%%%%%%%%%%%%%%
%%%%%%%%%%%%%%%%%%%%%%%%%%%%%%%%%%%%%%%%%%%%%%%%%%%%%%%%%%%%%%%%%%%%%%%%%%%%%%%%%%%

%\section{Spin texture and the $k.p$ model Hamiltonian}\label{sec:level4}

\textit{The $k.p$ model Hamiltonian and spin texture:} To probe the SC properties, we next construct a symmetry adapted low-energy \textbf{k.p} model Hamiltonian. The implementation of symmetry is easier in the product basis in the quantum numbers, namely the momentum, orbital, and spin. In presence of SOC, we consider the orbital states 
of $|l=1,m_l=\pm 1\rangle=(|p_x\rangle\pm i|p_y\rangle)\sqrt{2}$, and restrict ourselves to the total angular momentum $J=3/2$ manifold. We, thus write the single-particle Bloch state\cite{SharmaDas} of $|\psi_{{\bf k}ls}\rangle=Z_{\bf k}\otimes{W}_l\otimes{X}_s$, where orbital and spin spinors are ${W}_l=\left(|+\rangle ~ |-\rangle\right)^{T}$, and ${X}_s=\left(|\uparrow\rangle ~ |\downarrow\rangle\right)^{T}$, where $|\pm\rangle$ refer to $m_l=\pm 1$  and $|\uparrow,\downarrow\rangle$ denote $m_s=\pm 1/2$. Following Ref.~\onlinecite{SharmaDas}, we write $Z_{\bf k}=(z_{{\bf R}_1}({\bf k})~z_{{\bf R}_2}({\bf k}) ...)^{T}$ as a spinor of the plane waves:  $z_{\bf R}({\bf k})=e^{i{\bf k}\cdot{\bf R}}$, where ${\bf R}$ gives the unit cell position. The non-interacting Hamiltonian,  $H({\bf k})=\langle \psi_{{\bf k}ls}|H|\psi_{{\bf k}ls}\rangle$ can be expressed as

\begin{eqnarray}
H({\bf k})&=&\sum_{{\bf R}{\bf R'},\mu\nu}t_{\mu\nu}^{{\bf R}{\bf R'}}\mathcal{Z}_{{\bf R}{\bf R}'}({\bf k})\Gamma_{\mu\nu}=\sum_{\mu\nu}h_{\mu\nu}({\bf k})\Gamma_{\mu\nu}.
\end{eqnarray}
Here $\Gamma_{\mu\nu}=\sigma_{\mu}\otimes\tau_{\nu}$, $\tau_{\nu}$ and $\sigma_{\mu}$ being Pauli matrices in the orbital and spin bases, respectively, with $\mu,\nu = 0,x,y,z$. $\mu,\nu=0$ corresponds to $2\times 2$ identity matrices. Similarly, we define a $N\times N$ Bloch matrix $\mathcal{Z}_{{\bf R}{\bf R}'}({\bf k})=(Z_{\bf k}^{\dagger}Z_{\bf k})_{{\bf R}{\bf R}'}=z_{{\bf R}-{\bf R}'}({\bf k})$ , where $N$ is the number of nearest neighbors. $t^{{\bf R}{\bf R'}}_{\mu\nu}$ are the expansion coefficients or the tight-binding hopping tensors. %We define $h_{\mu\nu}({\bf k})=\sum_{{\bf R},{\bf R}'}t_{{\bf R}{\bf R}'}^{\mu\nu}\mathcal{Z}_{{\bf R}{\bf R}'}({\bf k})$. 

The symmetry operators in the spin and orbital spaces are given in S5 of SM \cite{supplemental}. Here we elaborate on the symmetry considerations in $\mathcal{Z}_{{\bf R}{\bf R}'}({\bf k})$. Owing to translational invariance, we set ${\bf R}=0$, and ${\bf R}'$ spans over six nearest neighbors in the triangular lattice formed by Te atoms, making $\mathcal{Z}({\bf k})=Z({\bf k})$ a six-component vector. Employing the $D_{3d}$ group, we decompose it into its irreducible representations (irreps) $Z=Z_{A_{1g}}\oplus Z_{A_{1u}}\oplus Z_{E_{g}}\oplus Z_{E_{1u}}$ at each ${\bf k}$. $A_{1g}$ and $E_{g}$ are even under inversion and contribute to the non-interacting Hamiltonian. $A_{1u}$ and $E_u$, being odd under inversion, though do not appear in non-interacting state, can appear in the broken symmetry superconducting state. Near $\Gamma$, the leading terms of $Z$, up to a normalization factor, are  given by (for full k-dependence see S6 of SM \cite{supplemental}):
\begin{eqnarray}
Z_{A_{1g}} &\approx& \alpha_0 +\alpha_1k_+k_- + \alpha_2 (k_+k_-)^2 + ... \nonumber\\ 
Z_{E_{g}^{(1/2)}}
&\approx& {\rm Re/Im}\left( \beta_1 k_+^2 + \beta_2 k_+^4 + ... \right), \nonumber\\
Z_{A_{1u}}&\approx&\gamma_1 k_+ + \gamma_2 k_+^3 + ...\nonumber\\
Z_{E_{u}^{(1/2)}}&\approx& {\rm Re/Im} (\delta_1 k_+^3 + \delta_2 k_+k_-^2 + \delta_2 k_+^3 + ...).
\end{eqnarray}

Here $k_{\pm}=k_x\pm ik_y$. The lowest-order terms represent $s$, $p$, $d$, $f$ - wave symmetries, while the higher-order terms are included to reproduce the warping term and spin textures in the outermost FSs. The coefficients $\alpha_i$, $\beta_i$, $\gamma_i$ and $\delta_i$ are the Taylor expansion coefficients of the corresponding $Z$ irreps (see SM).
\begin{eqnarray}
	%h_{00/zz} &=&  B_{0/z} + C_{0/z}k_+k_- + D_{0/z}(k_+k_-)^{2} \in Z_{A_{1g}} , \nonumber \\
	%h_{00/zz} &=& t_{00/zz}Z_{A_{1g}}, \nonumber\\ 
    h_{00} &=& t_{00}Z_{A_{1g}},~h_{zz} = t_{zz}Z_{A_{1g}}, \nonumber\\ 
    %h_{0x/0y}   &=& {\mathrm{Re/Im}\!\left(a_{1}k_{+}^{2}+A_{1}k_{-}^{4}\right)} \in Z_{E_{g}^{(1/2)}}, \nonumber \\
    %h_{0x/0y} &=& t_{0x/0y}{\mathrm{Re/Im}}\left(Z_{E_{g}^{(1/2)}}\right), \nonumber\\
    h_{0x} &=& t_{0x} Z_{E_{g}^{(1)}},~h_{0y} = t_{0y}Z_{E_{g}^{(2)}}, \nonumber\\
    %h_{xz/yz}  &=& {\mathrm{Re/Im}\!\left(a_{2}k_{+}^{2}+A_{2}k_{-}^{4}\right)} \in Z_{E_{g}^{(1/2)}}.% \nonumber \\
    %h_{xz/yz}  &=& t_{xz/yz}{\mathrm{Re/Im}}\left(Z_{E_{g}^{(1/2)}}\right).
    h_{xz}  &=& t_{xz}Z_{E_{g}^{(1)}},~h_{yz}  = t_{xz}Z_{E_{g}^{(2)}}.
\end{eqnarray}
Note, $h_{zz}$ corresponds to the Ising SOC, while $h_{xz/yz}$ includes the Rashba SOC effect that is reversed between the two orbitals, thereby maintaining the global inversion symmetry. By fitting the ${\bf k}\cdot{\bf p}$ model to the DFT FSs shown in Fig.~\ref{Fig:3}(a), we obtain the parameter values (in eV) to be $t_{00}(\alpha_{0}, \alpha_{1}, \alpha_{2}) =(0.18,\, 0,\,-13), ~ t_{zz}(\alpha_{0}, \alpha_{1}, \alpha_{2}) =(0.22,\,-4,\,13), ~ t_{0x}(\beta_{1}, \beta_2)=(0.8,\, 0.2), ~ t_{xz}(\beta_{1}, \beta_2)=(0.8,\, 3.6)$. 

Figures~\ref{Fig:3}(a)-(c) display the FS together with the corresponding spin textures obtained from DFT, while Figs.~\ref{Fig:3}(d)-(f) display the results from the ${\bf k}\cdot{\bf p}$ model. Results are shown for one Kramers partner; the other has opposite helicity, as required by inversion and time-reversal symmetries (see S7 of SM \cite{supplemental}). In particular, the Rashba like pattern within the Kramer degenerate pairs for the two inner sheets, arises from the real space point group symmetry of the Te atom, which belongs to the $C_{3v}$ group \cite{Zhang2014}. As is seen, the ${\bf k}\cdot{\bf p}$ model captures the 
key features of DFT FSs, in terms of no of sheets, detailed shapes and spin textures.  Some inaccuracies are observed for spin texture of innermost sheet. This sheet, however, contributes minimally to superconductivity, as discussed in the following.

\textit{Symmetry resolved Cooper pair:} 
Enforcing symmetries with fermion exchange is subtle. Following Ref.~\cite{SharmaDas}, we apply symmetries in the product basis for the Cooper pair wavefunction, defined as

\begin{equation}
\Psi({\bf k})=\mathsf{P}(\mathbb{Z}_{\bf k}\otimes\mathbb{W}\otimes\mathbb{X}),
\end{equation}
where $\mathbb{Z}$, $\mathbb{W}$ and $\mathbb{X}$ are the spatial, orbital, and spin sectors of the two-particle wavefunctions: 
%
%\begin{eqnarray}
%\mathbb{Z}_{\bf k}&=&\mathsf{P}_{\theta(Z)}\left[{Z}_{{\bf k}}\otimes {Z}_{-{\bf k}}\right], \nonumber\\
%\mathbb{W}_l&=&\mathsf{P}_{\theta(W)}\left[{W}_l\otimes {W}_l\right], \nonumber\\
%\mathbb{X}_s&=&\mathsf{P}_{\theta({X})}\left[{X}_s\otimes {X}_s\right]. 
%\end{eqnarray}
$\mathbb{Z}_{\bf k}=\mathsf{P}_{Z}\left[{Z}_{{\bf k}}\otimes {Z}_{-{\bf k}}\right]$, $\mathbb{W}_{ll'}=\mathsf{P}_{W}\left[{W}_l\otimes {W}_{l'}\right]$, 
$\mathbb{X}_{ss'}=\mathsf{P}_{X}\left[{X}_s\otimes {X}_{s'}\right]$. $\mathsf{P}=\mathsf{P}_{Z}\mathsf{P}_{W}\mathsf{P}_{X}$ ensures the total wavefunction must be odd under exchange. 

The next task is to impose the space-time and fermion exchange symmetries in all three wavefunction sectors. We decompose it into singlet/triplet ($\mathbb{X}^{(\mp)}$) components, where $\pm$ denotes the eigenvalues $\mathsf{P}_{X}$, as $\mathbb{X}_{ss'}=\mathbb{X}^{(-)}\oplus\mathbb{X}^{(+)}$.  Under the $D_{3d}$ point group, the orbital part decomposes into $\mathbb{W}_{ll'} = \mathbb{W}_{A_{1g}} \oplus \mathbb{W}_{A_{2g}} \oplus \mathbb{W}_{E_{g}}$, where  $\mathbb{W}_{A_{1g}}^{(+)} = \frac{1}{\sqrt{2}}\left(\left|+ -\right\rangle + \left|- +\right\rangle\right)$, $\mathbb{W}_{A_{2g}}^{(-)} = \frac{1}{\sqrt{2}}\left(\left|+ -\right\rangle - \left|- +\right\rangle\right)$, $\mathbb{W}_{E_{g}}^{(+)}= c_{1}\left|++\right\rangle + c_{2}\left|--\right\rangle$,
%    
%\begin{eqnarray}
%    \mathbb{W}_{A_{1g}}^{(+)} &=& \frac{1}{\sqrt{2}}\left(\left|+ -\right\rangle + \left|- +\right\rangle\right),\nonumber\\
%    \mathbb{W}_{A_{2g}}^{(-)} &=& \frac{1}{\sqrt{2}}\left(\left|+ -\right\rangle - \left|- +\right\rangle\right),\nonumber\\
%    \mathbb{W}_{E_{g}}^{(+)}&=& c_{1}\left|++\right\rangle + c_{2}\left|--\right\rangle,
%\end{eqnarray}
where $c_{1,2}$ are system-dependent parameters. It turns out the irreps are also eigen states of $\mathsf{P}_{W}$ with the corresponding eigenvalues $\pm$ denoted by superscript $(\pm)$. Finally, we consider the nearest-neighbor pairings, such that $\mathbb{Z}_{\bf k}=Z_{{\bf k}}=Z_{A_{1g}}\oplus Z_{A_{1u}}\oplus Z_{E_{g}}\oplus Z_{E_{1u}}$. The eigenvalues of $\mathsf{P}_Z$ coincides with the eigenvalues of the inversion symmetry, indicating that $Z^{(+)}_{A_{1g}}$, and $Z^{(+)}_{E_{g}}$ are even and $Z^{(-)}_{A_{1u}}$ and $Z^{(-)}_{E_{1u}}$ are odd under $\mathsf{P}_{Z}$. Although $Z^{(-)}_{A_{1u}}$ and $Z^{(-)}_{E_{1u}}$ are excluded from the non-interacting Hamiltonian by inversion symmetry, Cooper pairs can still host these odd-parity states if the total wavefunction remains antisymmetric.

Symmetry constraints reduce the allowed pairing channels. For example, even-parity sectors ($Z^{(+)}_{A_{1g}}$, $Z^{(+)}_{E_{g}}$) must combine with either odd-orbital ($\mathbb{W}_{A_{2g}}^{(-)}$) and even-spin ($\mathbb{X}^{(+)}$) or vice versa, while odd-parity sectors ($Z^{(-)}_{A_{1u}}$, $Z^{(-)}_{E_{1u}}$) require both orbital and spin parts to be simultaneously even or odd. The gap equation and energy minimization then select among these allowed channels. 
%Our numerical results yield  $Z^{(-)}_{A_{1u}}$ pairing on the smaller FS and $Z^{(-)}_{E_{1u}}$ on the larger one, both in the $\mathbb{W}_{A_{2g}}^{(-)}$ and spin-singlet  $\mathbb{X}^{(-)}$ channels. 
\\
%%%%%%%%%%%%%%%%%%%%%%%%%%%%%%%%%%%%%%%%%%%%%%%%%%%%%%%%%%%%%%

\textit{Numerical solution of spin-fluctuation mediated pairing symmetry-}
%\textit{Numerical results:}  
Starting with a genetic pairing potential, $V({\bf k},{{\bf k}'})$ we define a typical mean-field BCS gap equation \cite{Scalapino1986, Schrieffer1989, Sigrist1991, Takimoto2004, Mazin2008,Graser2009, Scalapino2012, Das2014} (see S8 of SM \cite{supplemental}),
 \begin{equation}
-\sum_{\nu',\textbf{k}'}V_{\nu,\nu'}(\textbf{k},\textbf{k}')\Delta_{\nu'}(\textbf{k}') =\lambda \Delta_{\nu}(\textbf{k}),
\label{eq-del_nu-2}
\end{equation} 
where $\Delta_{\nu}({\bf k})$ is the SC gap in the $\nu$th band, and $\lambda$ is the SC coupling strength. Eq.~\eqref{eq-del_nu-2} turns into an eigenvalue equation with $\lambda$ being the eigenvalue and $\Delta_{\nu}({\bf k})\propto \Psi_{l,s}({\bf k})$ being the eigenvectors. %In the weak-coupling limit, we restrict the momenta on the FSs. 

In the present context, conventional pairing mechanism such as electron-phonon coupling, with an attractive potential $V<0$, is weakened by the SOC and strong FS nesting. This makes the unconventional 
mechanism, such spin-fluctuation mediated with a repulsive potential ($V>0$) dominate. For $V>0$, a solution of $\Delta_{\nu}({\bf k})$ exists that changes sign between the FSs, i.e., ${\rm sgn}[\Delta_{\nu}({\bf k})]\ne {\rm sgn}[\Delta_{\nu'}({\bf k}')]$ promoted by the peaks in $V({\bf q})$ at ${\bf q}={\bf k}-{\bf k}'$. Strong FS nesting or the many-body spin-fluctuation interaction produces such dominant peaks, which stabilize an anisotropic pairing symmetry $\Delta_{\nu}({\bf k})$ with the largest eigenvalue $\lambda$. 

The Kanamori-Hubbard onsite interaction in the multi-band channel \cite{Hubbard1963,Hubbard1964,Kanamori,Georges2013,Ueda,DasDolui} produces an effective (repulsive) pairing interaction $V({\bf q})$. Decomposition of $V({\bf q})$ into spin, orbital, and spatial parts\cite{SharmaDas}, results in separate pairing interactions in the singlet and triplet $(\mp)$ spin irreps: 
\begin{equation}
\begin{aligned}
V^{(\mp)}(\textbf{q}) &= \pm \dfrac{1}{2}\left[\eta_{\mp}U_s \chi_s(\textbf{q})U_s \mp U_c\chi_c(\textbf{q})U_c \pm U_s \pm U_c)\right], \\
%\Gamma_{{+}}(\textbf{q}) &= -\dfrac{1}{2}\left[U_s\chi_s(\textbf{q})U_s + U_c\chi_c(\textbf{q})U_c-U_s-U_c)\right].
\end{aligned}
\end{equation}
where $\eta_{\mp}=1, 3$. Here $U_{s/c}$ and $\chi_{s/c}$ are the onsite Hubbard interactions and RPA susceptibilities in the spin ($s$) and charge ($c$) density-density fluctuation channels, respectively. $V^{\pm}$, $U_{s/c}$, and $\chi$ are matrices in the band basis, and the spatial dependence in $V^{\pm}({\bf q})$ arises from the RPA susceptibilities $\chi_{s/c}({\bf q})$. 

\begin{figure}[t]
\centering
\includegraphics[width=0.45\textwidth]{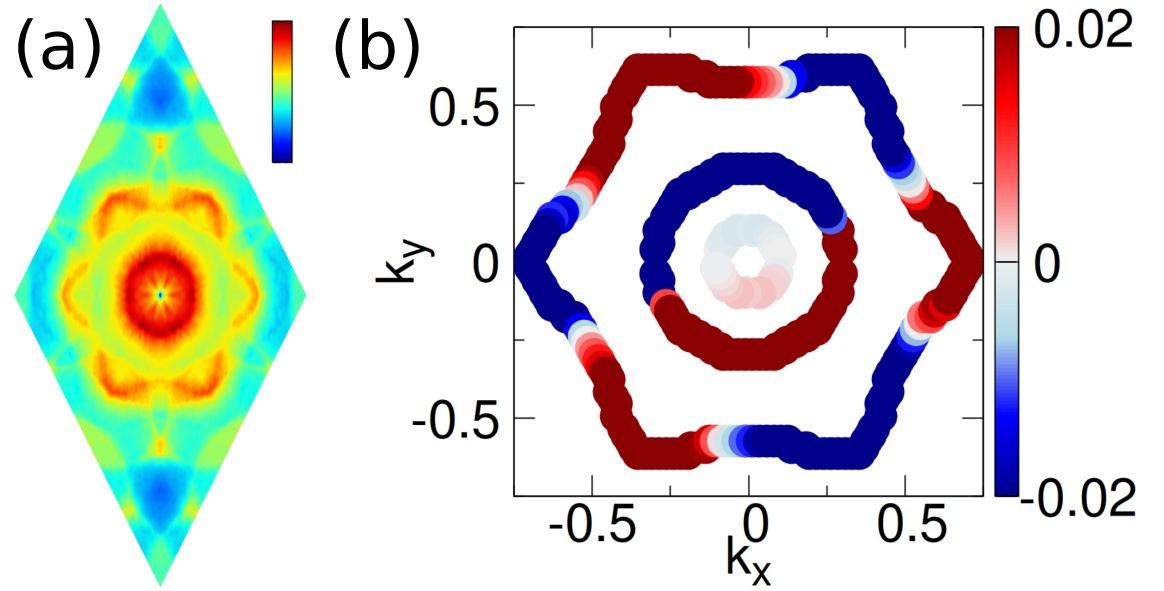}
\caption{\label{fig:pair} (a) RPA spin susceptibility $\chi_s({\bf q})$ and (b) pairing eigenvectors $\Delta_{\nu}({\bf k})$ visualized on the FS. The two concentric circular FSs are formed by degenerate bands 1 and 2, while the outer one is formed by degenerate bands 3 and 4. }
\end{figure}

The computed spin-fluctuation spectrum, $\chi_s({\bf q})$, is shown in Fig.~\ref{fig:pair} (a). It shows two dominant peaks, with a circular-shaped peak at a lower ${\bf q}$ value, and another at a higher ${\bf q}$ value, which are contributed dominantly by the inner and outer FSs (see Fig.~\ref{fig:pair}), respectively, with no discernible inter-band contribution. This is anticipated, as the two FSs have very different shapes, which weakens their inter-band nesting conditions. The smallest FS pocket's contribution to nesting is found to be very small. Calculation of the SC pairing symmetry and pairing eigenvalue, requires choice of intra and inter-band $U$'s. The results in Fig.~\ref{fig:pair}, are shown for representative choice of $1$ eV and 0.5 eV for VB1 and VB2, with 0.5 eV for
inter-band. Since the spin-fluctuation spectrum is dominated by the FS nestings, the specific values of interactions do not qualitatively change the conclusion as long as the non-interacting ground state remains intact, as checked by considering different choices of intra and inter-band $U$'s.

We solve Eq.~\eqref{eq-del_nu-2} numerically and plot the eigenvectors $\Delta_{\nu}({\bf k})$ as color maps on the FSs in Fig.~\ref{fig:pair}(b). We find pairings in both FSs occur in the odd-parity channels ($\mathbb{Z}^{(-)}_{A_u}$, and $\mathbb{Z}^{(-)}_{E_{u}}$) with spin singlet $\mathbb{X}^{(-)}$, and orbital-singlet $\mathbb{W}_{A_{2g}}^{(-)}$.  The leading term of $Z_{A_{1u}}$ has the $p_{x/y}$ wave symmetry on the inner FSs, while the outer FS  is warped by the higher order term i.e.,  ${\rm Re/Im}(k_+^3)$, giving $f_{x(x^2-3y^2)}$, $f_{y(y^2-3x^2)}$ wave symmetry. Notably, the pairing on outer most FS 
follows the Ising spin textures $\langle \sigma_z\rangle$ (Fig.~3(c,f)), while that in inner FS sheet follows the Rashba spin texture $\langle \sigma_{x,y}\rangle$ (Figs. 3(a-e)).
Note, the Rashba pairing is helical due to the strong in-plane SOC.
This suggests both Ising SOC ($h_{zz}$) and 
Rashba SOC ($h_{xz}, h_{yz})$ drive pairing between opposite spins and orbitals at $\pm{\bf k}$ in the corresponding electronic structure of 2D IrTe$_2$, albeit in two different FSs. 
2D IrTe$_2$ thus presents a rare case of band-selective co-existence of Ising and Rashba superconductivity. 
 
\textit{Conclusion:} We uncover in strain-stabilized monolayer IrTe$_2$ a band-selective coexistence of Rashba (in-plane) and Ising (out-of-plane) spin texture induced Cooper pairings.  Both pairings are odd under fermion exchange in spin, orbital, and momentum channels, and protected from mixing by inversion symmetry and distinct irreducible representations. This unique situation overcomes the parity-mixing and singlet-triplet mixing limitations of inversion-broken Rashba systems. This in turn enables effective spin filtering, large and anisotropic $H_{c2}$, and spin-sensitive Josephson responses that selectively probe in-plane versus out-of-plane Cooper pairs. Our investigation on 2D IrTe$_2$ thus open up the way for symmetry-and-spin-texture engineering as a general strategy to realize multichannel odd pairing in 2D TMDs. This may be tuned via strain, carrier density, and electric field.
%This opens up a vast landscape of materials and devices, which is expected to facilitate experiments and applications in SC logic, quantum interconnects, and spintronics.

\textit{Acknowledgments-} K.D. thanks the Council of Scientific and Industrial Research (CSIR) for support through a fellowship (File No. 09/080(1178)/2020-EMR-I). I.D. thanks the Science and Engineering Research Board (SERB) India (Project No. CRG/2021/003024) and the Technical Research Center, Department of Science, and Technology for support. R.S. acknowledges the Science and Engineering Research Board (SERB), Government of India, for providing the National Post Doctoral Fellowship (NPDF) with Grant No. PDF/2021/000546. T.D. acknowledges Anusandhan National Research Foundation (ANRF), India for the research funding under Core Research Grant (CRG) (Grant No.: CRG/2022/003412). T.S.D. acknowledges JC Bose National Fellowship (Grant No. JCB/2020/000004) for funding during the work. 

\textit{Data availability-} The data supporting this study’s findings are available within the article.

\nocite{Matsumoto1999}
\nocite{PhysRevB.62.R16219}
\nocite{PhysRevLett.53.2571}
\nocite{TOGO20151}
\nocite{doi:10.7566/JPSJ.92.012001}
\nocite{Bucher2011}
\nocite{Ramadevi_Dubey_2019}

%\bibliography{manuscript}

%apsrev4-2.bst 2019-01-14 (MD) hand-edited version of apsrev4-1.bst
%Control: key (0)
%Control: author (8) initials jnrlst
%Control: editor formatted (1) identically to author
%Control: production of article title (0) allowed
%Control: page (0) single
%Control: year (1) truncated
%Control: production of eprint (0) enabled
%

%%%%%%%%%%%%%%%%%%%%%%%%%%%%%%%%%%%%%%%%%%%%%%%%%%%%%%%%%%%%%%%%%%%%%%%%
%%% SUPPLEMENTAL MATERIAL %%%
%%%%%%%%%%%%%%%%%%%%%%%%%%%%%%%%%%%%%%%%%%%%%%%%%%%%%%%%%%%%%%%%%%%%%%%%
\clearpage
\onecolumngrid

\begin{center}
	{\Large \textbf{Supplemental Information}} \\
	\vspace{0.6 cm}
	{\large\textbf{Coexistence of Rashba and Ising Spin-Singlet Pairings in Two-Dimensional IrTe$_{2}$}}\\
	\vspace{0.3 cm}
	{{Kunal Dutta$^{1}$}, {Rajesh O. Sharma$^{2,3}$}, {Shreya Das$^{4,5}$}, {Indra Dasgupta$^{1}$}, {Tanmoy Das$^{2}$}, and {Tanusri Saha-Dasgupta$^{4}$} }\\
	\vspace{0.4 cm}
	{$^{1}$School of Physical Sciences, Indian Association for the Cultivation of Science, 2A and 2B Raja S.C. Mullick Road, Jadavpur, Kolkata 700032, India.\\
		$^{2}$ Department of Physics, Indian Institute of Science, Bangalore 560012, India.\\
		$^{3}$ Department of Science and Humanities (Physics), Dr. B. R. Ambedkar Institute of Technology, Dollygunj, Sri Vijaya Puram 744103, Andaman and Nicobar Islands, India.\\
		$^{4}$ Department of Condensed Matter and Materials Physics, S. N. Bose National Centre for Basic Sciences, JD Block, Sector III, Salt Lake, Kolkata, West Bengal 700106, India.\\
		$^{5}$ Masaryk University, Kamenice 753/5, 625 00 Brno, Czech Republic.
	}
\end{center}

% Reset numbering for supplement
\setcounter{section}{0}
\setcounter{figure}{0}
\setcounter{table}{0}

\renewcommand{\thesection}{S\arabic{section}}
\renewcommand{\thefigure}{S\arabic{figure}}
\renewcommand{\thetable}{S\arabic{table}}

%\preprint{APS/123-QED}
%\title{Coexistence of Rashba and Ising Spin-Singlet Pairings in Two-Dimensional IrTe$_{2}$}
%\author{Kunal Dutta$^{1}$} 
%\author{Rajesh O. Sharma$^{2,3}$}
%\author{Shreya Das$^{4,5}$}
%\author{Indra Dasgupta$^{1}$}  \email{sspid@iacs.res.in}   
%\author{Tanmoy Das$^{2}$} \email{tnmydas@iisc.ac.in}
%\author{Tanusri Saha-Dasgupta$^{4}$} \email{t.sahadasgupta@gmail.com}
%\affiliation{$^{1}$School of Physical Sciences, Indian Association for the Cultivation of Science, 2A and 2B Raja S.C. Mullick Road, Jadavpur, Kolkata 700032, India.\\
%	$^{2}$ Department of Physics, Indian Institute of Science, Bangalore 560012, India.\\
%	$^{3}$ Department of Science and Humanities (Physics), Dr. B. R. Ambedkar Institute of Technology, Dollygunj, Sri Vijaya Puram 744103, Andaman and Nicobar Islands, India.\\
%	$^{4}$ Department of Condensed Matter and Materials Physics, S. N. Bose National Centre for Basic Sciences, JD Block, Sector III, Salt Lake, Kolkata, West Bengal 700106, India.\\
%	$^{5}$ Masaryk University, Kamenice 753/5, 625 00 Brno, Czech Republic.
%}
%\date{\today}
%\maketitle
%\vspace{0.8 cm}

\section{Computational Details}
The first-principles calculations presented in this work were performed using the Vienna \emph{ab initio} Simulation Package (VASP)~\cite{PhysRevB.47.558,PhysRevB.54.11169} within the framework of density functional theory (DFT). The electron-ion interaction was described using the projector augmented-wave (PAW) method~\cite{PhysRevB.50.17953}, and exchange-correlation effects were treated using the Perdew-Burke-Ernzerhof (PBE) generalized gradient approximation (GGA)~\cite{PhysRevLett.77.3865}. A plane-wave energy cutoff of 600~eV was employed to ensure accurate convergence. For self-consistent field (SCF) calculations, the Brillouin zone was sampled using a $12 \times 12 \times 1$ Monkhorst-Pack $k$-point mesh.

For an effective low-energy description, a few-band tight-binding Hamiltonian and the corresponding Wannier functions were derived from the converged DFT electronic structure using the $N$th-order muffin-tin orbital (NMTO) downfolding technique~\cite{PhysRevB.62.R16219}. The self-consistent potential parameters required for the NMTO calculations were obtained from linear muffin-tin orbital (LMTO) calculations~\cite{PhysRevLett.53.2571}. The consistency between the plane-wave and muffin-tin orbital basis sets was carefully validated by comparing the electronic band structures and density of states (DOS) obtained from both approaches.

Lattice dynamical properties were investigated using the finite displacement method as implemented in VASP. The force constants were calculated using a $2 \times 2 \times 1$ supercell. Phonon dispersions were subsequently obtained by Fourier interpolation of the real-space force constants using the \textsc{Phonopy} package~\cite{TOGO20151,doi:10.7566/JPSJ.92.012001}, and plotted along the high-symmetry paths of the Brillouin zone.

Finally, the thermodynamic stability of the system was assessed via \emph{ab initio} molecular dynamics (AIMD) simulations performed within the canonical ($NVT$) ensemble. The temperature was controlled using a Nos\'e-Hoover thermostat~\cite{Bucher2011} at 300~K. Each AIMD simulation was carried out for 5.0~ps with a time step of 1~fs.

%%%%%%%%%%%%%%%%%%%%%%%%%%%%%%%%%%%%%%%%%%%%%%%%%%%%%%%%%%%%%%%%%%%%%%%%%%%%%%%%%%
%\begin{figure}[h]
%	\includegraphics[height= 50 mm, width=150 mm,keepaspectratio]{SFig1.pdf}
%	\caption{ (a) and (b) are schematic representation of type-I and type-II Ising superconductivity, respectively.}
%\end{figure}
%%%%%%%%%%%%%%%%%%%%%%%%%%%%%%%%%%%%%%%%%%%%%%%%%%%%%%%%%%%%%%%%%%%%%%%%%%%%%%%%%%
\section{Structural Stability}
\begin{figure}[h]
	\includegraphics[height= 50 mm, width=150 mm,keepaspectratio]{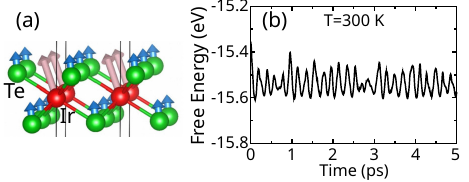}
	\caption{ (a) Displacement vector corresponding to the imaginary
		(negative) phonon mode.  (b) Variation of the free energy as a function of time \(t\) at \(T = 300\ \mathrm{K}\) for IrTe$_2$ monolayer respectively.}
	\label{SFig:1}
\end{figure}

At room temperature, bulk IrTe$_2$ crystallizes in the centrosymmetric space group $P\bar{3}m1$, with lattice parameters $a = 3.928$~\AA{} and $c = 5.394$~\AA{} \cite{Matsumoto1999}. A monolayer cleaved from this bulk structure is found to be dynamically unstable, as evidenced by the appearance of imaginary (negative) phonon frequencies in its vibrational spectrum and the displacement vector corresponding to the imaginary (negative) phonon mode is shown in Fig.~\ref{SFig:1}(a). A detailed analysis of the unstable phonon modes indicates that the application of biaxial strain can effectively stabilize the monolayer. Upon applying biaxial strain and fully relaxing the structure, the optimized monolayer attains lattice parameters $a = 3.968(8)$~\AA{} and $c = 17.682(7)$~\AA{}. Importantly, the strained monolayer retains the same space group symmetry, $P\bar{3}m1$, as the bulk phase. Consequently, inversion symmetry is preserved, consistent with the centrosymmetric nature of the parent bulk structure. In the relaxed monolayer, the Ir atom occupies the Wyckoff position $1b$ with site symmetry $D_{3d}$, while the Te atoms reside at the Wyckoff position $2d$ with site symmetry $C_{3v}$. The corresponding atomic coordinates within the unit cell are summarized in Table~\ref{stab:table1}. Finally, the thermodynamic stability of the system was evaluated using \emph{ab initio} molecular dynamics (AIMD) simulations, as shown in Fig.~\ref{SFig:1}(b).

%%%%%%%%%%%%%%%%%%%%%%%%%%%%%%%%%%%%%%%%%%%%%%%%%%%%%%%%%%%%%%%%%%%%%%%%%%%%%%%%%%
%%%%%%%%%%%%%%%%%%%%%%%%%%%%%%%%%%%%%%%%%%%%%%%%%%%%%%%%%%%%%%%%%%%%%%%%%%%%%%%%%%

%%%%%%%%%%%%%%%%%%%%%%%%%%%%%%%%%%%%%%%%%%%%%%%%%%%%%%%%%%%%%%%%%%%%%%%%%%%%%%%%%%
\begin{table}[h]
	\caption{ Crystal Structure and Wyckoff Positions of monolayer IrTe$_2$. }
	\begin{ruledtabular}
		\begin{tabular}{lccc}
			\multicolumn{4}{c}{IrTe$_2$}\tabularnewline
			\hline 
			Space group & a $\mathring{\left(A\right)}$ & b $\mathring{\left(A\right)}$ & c $\mathring{\left(A\right)}$\tabularnewline
			
			$P\bar{3}m1$ & 3.9688 & 3.9688 & 17.68275\tabularnewline
			
			\hline 
			Atom & x & y & z\tabularnewline
			\hline 
			Ir (1b) & 0.0 & 0.0 & 0.5 \tabularnewline
			Te  (2b) & 0.666667  & 0.333332 & 0.579238 
		\end{tabular} 
	\end{ruledtabular}
	\label{stab:table1}
\end{table}
\section{On-Site Matrix Elements}
To construct an effective low-energy description, we consider a few-band tight-binding Hamiltonian composed of Te-$p_{x}$, $p_{y}$, $p_{z}$ and Ir-$d_{xy}$, $d_{yz}$, and $d_{xz}$ orbitals. The tight-binding Hamiltonian and the corresponding Wannier functions were derived from the converged DFT electronic structure using the $N$th-order muffin-tin orbital (NMTO) downfolding technique~\cite{PhysRevB.62.R16219}.

The on-site matrix elements in units of eV for the Ir-$d_{xy}$, $d_{yz}$, and $d_{xz}$ orbitals, as well as for the Te-$p_{y}$, $p_{z}$, and $p_{x}$ orbitals, are presented below.

\begin{equation}
	H^{\text{Ir}}_{\text{on-site}} =
	\begin{pmatrix}
		\langle d_{xy}|H|d_{xy}\rangle & \langle d_{xy}|H|d_{yz}\rangle & \langle d_{xy}|H|d_{xz}\rangle \\
		\langle d_{yz}|H|d_{xy}\rangle & \langle d_{yz}|H|d_{yz}\rangle & \langle d_{yz}|H|d_{xz}\rangle \\
		\langle d_{xz}|H|d_{xy}\rangle & \langle d_{xz}|H|d_{yz}\rangle & \langle d_{xz}|H|d_{xz}\rangle
	\end{pmatrix}
	=
	\begin{pmatrix}
		-3.3449 & 0.4146 & 0.0000 \\
		0.4146  & -3.0775 & 0.0000 \\
		0.0000  & 0.0000  & -3.2274
	\end{pmatrix}. \nonumber
\end{equation}

\begin{equation}
	H^{\text{Te}}_{\text{on-site}} =
	\begin{pmatrix}
		\langle p_{y}|H|p_{y}\rangle & \langle p_{y}|H|p_{z}\rangle & \langle p_{y}|H|p_{x}\rangle \\
		\langle p_{z}|H|p_{y}\rangle & \langle p_{z}|H|p_{z}\rangle & \langle p_{z}|H|p_{x}\rangle \\
		\langle p_{x}|H|p_{y}\rangle & \langle p_{x}|H|p_{z}\rangle & \langle p_{x}|H|p_{x}\rangle
	\end{pmatrix}
	=
	\begin{pmatrix}
		-3.1777 & 0.0000 & 0.0000 \\
		0.0000  & -2.1016 & 0.0920 \\
		0.0000  & 0.0920  & -2.5256
	\end{pmatrix}. \nonumber
\end{equation}

%%%%%%%%%%%%%%%%%%%%%%%%%%%%%%%%%%%%%%%%%%%%%%%%%%%%%%%%%%%%%%%%%%%%%%%%%%%%%%%%%%
\section{Electronic structure}
\begin{figure*}[h]
	\includegraphics[height= 80 mm, width=178 mm,keepaspectratio]{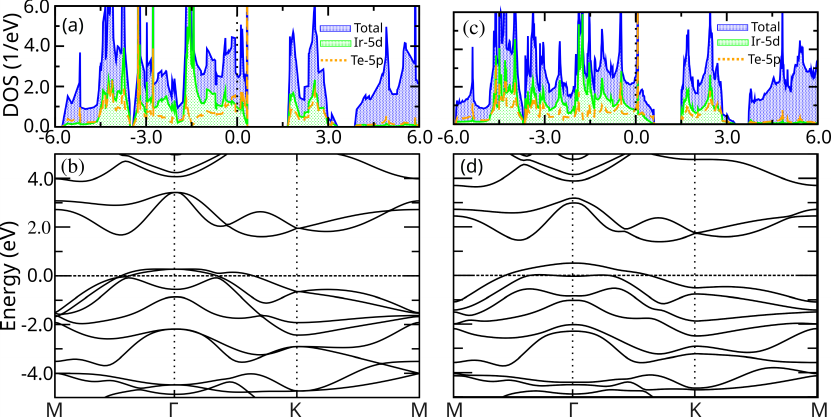}
	\caption{The density of states (DOS) and electronic band structure of an IrTe$_2$ monolayer with and without SOC. (a) The total and projected DOS for the Ir-5$d$ and Te-5$p$ orbitals without SOC. (b) The corresponding band structures along high-symmetry points of the Brillouin zone without SOC. (c) The total and projected DOS for the Ir-5$d$ and Te-5$p$ orbitals with SOC. (d) The corresponding band structures along high-symmetry points of the Brillouin zone with SOC. The band structures are plotted along the M-$\Gamma$-K-M path of the Brillouin zone. The Fermi level is set to zero on the energy axis. }
	\label{sfig:2}
\end{figure*}

The density of states (DOS) of the stable monolayer IrTe$_2$ in the absence of spin-orbit coupling (SOC), obtained from first-principles DFT calculations, is shown in Fig.~\ref{sfig:2}(a). The non-spin-polarized total and orbital-projected DOS reveal that monolayer IrTe$_2$ exhibits metallic behavior. In the vicinity of the Fermi energy $E_F$, the valence states are predominantly contributed by Te-$p$ orbitals. As illustrated in Fig.~\ref{sfig:2}(a), the Ir-$d$ states are broadly distributed below $E_F$, with only a small portion remaining unoccupied. These relatively delocalized Ir-$d$ bands overlap in energy with the bonding and antibonding states formed by the Te-$p$ orbitals. The metallic nature of monolayer IrTe$_2$ is further corroborated by the electronic band structure calculated without SOC, as shown in Fig.~\ref{sfig:2}(b). The total and orbital-projected DOS including SOC for both Ir-$d$ and Te-$p$ states are presented in Fig.~\ref{sfig:2}(c). The inclusion of SOC does not lead to any significant modification of the overall DOS compared to the SOC-free case. The corresponding band structure including SOC is displayed in Fig.~\ref{sfig:2}(d). A salient feature of the SOC-included band structure is the additional band splitting induced by SOC. In particular, at the $\Gamma$ point, the $p_x$ and $p_y$ bands are no longer degenerate. Nevertheless, Kramers degeneracy is preserved throughout the Brillouin zone as a consequence of the combined presence of time-reversal and inversion symmetries.
%%%%%%%%%%%%%%%%%%%%%%%%%%%%%%%%%%%%%%%%%%%%%%%%%%%%%%%%%%%%%%%%%%%%%%%%%%%%%%%%%%
\section{Symmetry operations of the \texorpdfstring{$D_{3d}$}{D3d} point group}
%%%%%%%%%%%%%%%%%%%%%%%%%%%%%%%%%%%%%%%%%%%%%%%%%%%%%%%%%%%%%%%%%%%%%%%%%%%%%%%%%%%%%%%%%%
The transformation rules for the crystal momentum $\mathbf{k}$, as well as for the Pauli matrices $\sigma$ and $\tau$  in the spin and orbital subspaces, respectively, under the considered point-group symmetry operations \cite{PhysRevLett.123.126402} are summarized in Table \ref{stab:table2}.

\begin{table*}[h]
	\centering
	\caption{Symmetry operations of $D_{3d}$ at the $\Gamma$ point. }
	\begin{ruledtabular}
		\begin{tabular}{c|c|c|c}
			Symmetry Operation & $(k_x, k_y)$ & $(\sigma_x, \sigma_y, \sigma_z)$ & $(\tau_x, \tau_y, \tau_z)$ \tabularnewline
			\hline 
			$T = -i\sigma_y \otimes \tau_x K$ & $(-k_x, -k_y)$ & $(-\sigma_x, -\sigma_y, -\sigma_z)$ & $(\tau_x, \tau_y, -\tau_z)$ \tabularnewline
			$M_{x} = -i\sigma_x \otimes (-\tau_x)$ & $(-k_x, k_y)$ & $(\sigma_x, \sigma_y, \sigma_z)$ & $(\tau_x, -\tau_y, -\tau_z)$ \tabularnewline
			$P = \sigma_0 \otimes \tau_0$ & $(-k_x, -k_y)$ & $(\sigma_x, \sigma_y, \sigma_z)$ & $(\tau_x, \tau_y, \tau_z)$ \tabularnewline
			$C_{3z} = e^{-i\frac{\pi}{3} \sigma_{z}} \otimes e^{-i\frac{2\pi}{3} \tau_{z}}$ & 
			$\left(-\frac{1}{2}k_x + \frac{\sqrt{3}}{2}k_y,\; -\frac{\sqrt{3}}{2}k_x - \frac{1}{2}k_y\right)$ & 
			$\left(\frac{1}{2}\sigma_x + \frac{\sqrt{3}}{2}\sigma_y,\; -\frac{\sqrt{3}}{2}\sigma_x + \frac{1}{2}\sigma_y,\; \sigma_z\right)$ & 
			$\left(-\frac{1}{2}\tau_x + \frac{\sqrt{3}}{2}\tau_y,\; -\frac{\sqrt{3}}{2}\tau_x - \frac{1}{2}\tau_y,\; \tau_z\right)$ \tabularnewline
			\label{stab:table2}
		\end{tabular}
	\end{ruledtabular}
\end{table*}

%%%%%%%%%%%%%%%%%%%%%%%%%%%%%%%%%%%%%%%%%%%%%%%%%%%%%%%%%%%%%%%%%%%%%%%%%%%%%%%%%%

\section{Irreducible representation of $Z({\bf k})$, $\mathbb{W}_l$}

\subsection{ Matrix representation of symmetry operations in the orbital basis }

A reducible representation $\Gamma$ of a point group $D_{3d}$ 
can be expressed as a direct sum of the irreducible representations 
$\Gamma^{\alpha}$ of $G$, in which each irreducible representation appears 
with a certain multiplicity $m_{\alpha}$. This can be written as \cite{Ramadevi_Dubey_2019}
\begin{equation}
	\Gamma = \bigoplus_{\alpha=1}^{r} m_{\alpha}\, \Gamma^{\alpha}. 
	\label{Eq:A1}
\end{equation}

The multiplicity $m_{\alpha}$ of each irreducible representation $\Gamma^{\alpha}$ 
is obtained using the standard formula \cite{Ramadevi_Dubey_2019}
\begin{equation}
	m_{\alpha} = \frac{1}{|G|} \sum_{g \in G} 
	\chi^{\Gamma}(g)\, \overline{\chi^{\Gamma^{\alpha}}(g)},
	\label{Eq:A2}
\end{equation}
where $|G|$ denotes the order of the group, $\chi^{\Gamma}(g)$ is the character 
of the reducible representation $\Gamma$ corresponding to the group element $g$, 
and $\chi^{\Gamma^{\alpha}}(g)$ is the character of the irreducible representation 
$\Gamma^{\alpha}$.

As we have chosen the orbital basis for the two-particle system as $\left|++\right\rangle$, $\left|+-\right\rangle$, $\left|-+\right\rangle$, and $\left|--\right\rangle$, the symmetry operations of the point group $D_{3d}$ in this basis can be represented in the following matrix form:

%\blue{Once you defined the basis you do not need to repeat them in the table below}
%As we have chosen the the orbital basis for two particle as $\left|++\right\rangle $ , $\left|+-\right\rangle $, $\left|-+\right\rangle $ , $\left|--\right\rangle $ . The symmetry operations under the Point group $D_{3d}$ in this particular basis in has the matrix forms as:
%%%%%%%%%%%%%%%%%%%%%%%%%%%%%%%%%%%%%%%%%%%%%%%%%%%%%%%%%%%%%%%%%%%%%%%%%%%%%%%%%%%%%%%%%%

\begin{eqnarray*}
	E=\left(\begin{array}{cccc}
		1 & 0 & 0 & 0\\
		0 & 1 & 0 & 0\\
		0 & 0 & 1 & 0\\
		0 & 0 & 0 & 1
	\end{array}\right),  & \quad I=\left(\begin{array}{ccccc}
		1 & 0 & 0 & 0\\
		0 & 1 & 0 & 0\\
		0 & 0 & 1 & 0\\
		0 & 0 & 0 & 1 
	\end{array}\right),   \quad C_{3}=\left(\begin{array}{cccc}
		e^{-i\frac{4\pi}{3}} & 0 & 0 & 0\\
		0 & 1 & 0 & 0\\
		0 & 0 & 1 & 0\\
		0 & 0 & 0 & e^{i\frac{4\pi}{3}}
	\end{array}\right),  & \quad C_{3}^{2}=\left(\begin{array}{ccccc}
		e^{-i\frac{2\pi}{3}} & 0 & 0 & 0\\
		0 & 1 & 0 & 0\\
		0 & 0 & 1 & 0\\
		0 & 0 & 0 & e^{i\frac{2\pi}{3}} \end{array}\right) \\
\end{eqnarray*}

\begin{eqnarray*}
	C_{2}^{\prime}=\left(\begin{array}{cccc}
		0 & 0 & 0 & 1\\
		0 & 0 & 1 & 0\\
		0 & 1 & 0 & 0\\
		1 & 0 & 0 & 0
	\end{array}\right), &  \quad C_{2}^{\prime}C_{3}=\left(\begin{array}{ccccc}
		0 & 0 & 0 & e^{-i\frac{4\pi}{3}}\\
		0 & 0 & 1 & 0\\
		0 & 1 & 0 & 0\\
		e^{i\frac{4\pi}{3}} & 0 & 0 & 0 
	\end{array}\right), & \quad C_{2}^{\prime}C^2_{3}=\left(\begin{array}{cccc}
		0 & 0 & 0 & e^{-i\frac{2\pi}{3}}\\
		0 & 0 & 1 & 0\\
		0 & 1 & 0 & 0\\
		e^{i\frac{2\pi}{3}} & 0 & 0 & 0
	\end{array}\right),   \quad S_{6}=\left(\begin{array}{ccccc}
		e^{i\frac{2\pi}{3}} & 0 & 0 & 0\\
		0 & 1 & 0 & 0\\
		0 & 0 & 1 & 0\\
		0 & 0 & 0 & e^{-i\frac{2\pi}{3}} \end{array}\right) \\
\end{eqnarray*}

\begin{eqnarray*}
	S_{6}^{2}=\left(\begin{array}{cccc}
		e^{i\frac{4\pi}{3}} & 0 & 0 & 0\\
		0 & 1 & 0 & 0\\
		0 & 0 & 1 & 0\\
		0 & 0 & 0 & e^{-i\frac{4\pi}{3}}
	\end{array}\right), &   \quad \sigma_{d}=\left(\begin{array}{ccccc}
		0 & 0 & 0 & 1\\
		0 & 0 & 1 & 0\\
		0 & 1 & 0 & 0\\
		1 & 0 & 0 & 0 
	\end{array}\right), \quad \sigma_{d}C_{3}=\left(\begin{array}{cccc}
		0 & 0 & 0 & e^{i\frac{2\pi}{3}}\\
		0 & 0 & 1 & 0\\
		0 & 1 & 0 & 0\\
		e^{-i\frac{2\pi}{3}} & 0 & 0 & 0
	\end{array}\right),  & \quad  \sigma_{d}C^2_{3}=\left(\begin{array}{ccccc}
		0 & 0 & 0 & e^{-i\frac{2\pi}{3}}\\
		0 & 0 & 1 & 0\\
		0 & 1 & 0 & 0\\
		e^{i\frac{2\pi}{3}} & 0 & 0 & 0 \end{array}\right) \\
\end{eqnarray*}

The characters of $\Gamma$ are then computed as:
\begin{align}
	\chi^{\Gamma}(E) &= \chi^{\Gamma}(I) = 4, \nonumber \\
	\chi^{\Gamma}(C_{3}) &= \chi^{\Gamma}(C_{3}^{2}) = 1, \nonumber \\
	\chi^{\Gamma}(S_{6}) &= \chi^{\Gamma}(S_{6}^{2}) = 1, \nonumber \\
	\chi^{\Gamma}(C_{2}^{\prime}) &= \chi^{\Gamma}(C_{2}^{\prime}C_{3}) 
	= \chi^{\Gamma}(C_{2}^{\prime}C_{3}^{2}) = 0, \nonumber \\
	\chi^{\Gamma}(\sigma_{d}) &= \chi^{\Gamma}(\sigma_{d}C_{3}) 
	= \chi^{\Gamma}(\sigma_{d}C_{3}^{2}) = 0. \nonumber
\end{align}

The decomposition of $\Gamma$ into its irreducible components can be obtained by  evaluating the multiplicities using Eq.~(\ref{Eq:A2}). We find that $\Gamma$ reduces to the direct sum
\begin{equation}
	\Gamma = A_{1g} \oplus A_{2g} \oplus E_{g}. \nonumber
\end{equation}

The projection operator $P_{\Gamma^{\alpha}}$ acts on a reducible representation  $\Gamma$, allowing it to be decomposed into its irreducible components.  It is defined as
\begin{equation}
	P_{\Gamma^{\alpha}} = \frac{\ell_{\alpha}}{|G|} 
	\sum_{g \in G} \chi^{\alpha}(g)\, \Gamma(g),
	\label{eq:projection_operator}
\end{equation}
where $\ell_{\alpha}$ is the dimension of the irreducible representation  $\Gamma^{\alpha}$, $|G|$ is the order of the group, $\chi^{\alpha}(g)$ is  the character of $\Gamma^{\alpha}$, and $\Gamma(g)$ is the matrix  representation of the group element $g$. Applying the projection operator yields the basis states that transform  according to the irreducible representations. In the present case, the  projected basis states are
\begin{itemize}
	\item $A_{1g}$ (even): 
	\[
	\mathbb{W}_{A_{1g}}^{(+)} = 
	\tfrac{1}{\sqrt{2}}\left(\,\left|+-\right\rangle + \left|-+\right\rangle\right),
	\]
	
	\item $A_{2g}$ (odd): 
	\[
	\mathbb{W}_{A_{2g}}^{(-)} = 
	\tfrac{1}{\sqrt{2}}\left(\,\left|+-\right\rangle - \left|-+\right\rangle\right),
	\]
	
	\item $E_{g}$ (even): 
	\[
	\mathbb{W}_{E_{g}}^{(+)} = 
	c_{1}\left|++\right\rangle + c_{2}\left|--\right\rangle,
	\]
	where $c_{1}$ and $c_{2}$ are appropriate normalization constants.
\end{itemize}

\subsection{ Matrix representation of symmetry operations in the plane wave basis }

\begin{figure}[h]
	\centering
	\includegraphics[width=80mm,height=53mm,keepaspectratio]{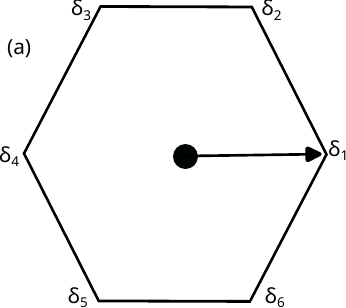}\caption{\label{fig:figure4} The Brillouin Zone of Triangular lattice.}
	%\blue{Do not include kx,ky in the delta.}}
	\end{figure}
	
	Here the basis set is 
	\begin{eqnarray*}
\left|\psi\right\rangle  & = & \left(\begin{array}{c}
	e^{i\vec{k}.\delta_{1}} \quad e^{i\vec{k}.\delta_{2}} \quad e^{i\vec{k}.\delta_{3}} \quad e^{i\vec{k}.\delta_{4}} \quad e^{i\vec{k}.\delta_{5}} \quad e^{i\vec{k}.\delta_{6}}
\end{array}\right) {^{T}}
\end{eqnarray*}

The symmetry operations of the point group $D_{3d}$ in this plane wave basis can be represented in the following matrix form:

\begin{eqnarray*}
I=\left(\begin{array}{cccccc}
	0 & 0 & 0 & 1 & 0 & 0\\
	0 & 0 & 0 & 0 & 1 & 0\\
	0 & 0 & 0 & 0 & 0 & 1\\
	1 & 0 & 0 & 0 & 0 & 0\\
	0 & 1 & 0 & 0 & 0 & 0\\
	0 & 0 & 1 & 0 & 0 & 0
\end{array}\right) ,  \quad C_{3}=\left(\begin{array}{cccccc}
	0 & 0 & 1 & 0 & 0 & 0\\
	0 & 0 & 0 & 1 & 0 & 0\\
	0 & 0 & 0 & 0 & 1 & 0\\
	0 & 0 & 0 & 0 & 0 & 1\\
	1 & 0 & 0 & 0 & 0 & 0\\
	0 & 1 & 0 & 0 & 0 & 0
\end{array}\right), \quad C_{3}^{2}=\left(\begin{array}{cccccc}
	0 & 0 & 0 & 0 & 1 & 0\\
	0 & 0 & 0 & 0 & 0 & 1\\
	1 & 0 & 0 & 0 & 0 & 0\\
	0 & 1 & 0 & 0 & 0 & 0\\
	0 & 0 & 1 & 0 & 0 & 0\\
	0 & 0 & 0 & 1 & 0 & 0
\end{array}\right),  \quad S_{6}=\left(\begin{array}{cccccc}
	0 & 1 & 0 & 0 & 0 & 0\\
	0 & 0 & 1 & 0 & 0 & 0\\
	0 & 0 & 0 & 1 & 0 & 0\\
	0 & 0 & 0 & 0 & 1 & 0\\
	0 & 0 & 0 & 0 & 0 & 1\\
	1 & 0 & 0 & 0 & 0 & 0
\end{array}\right)
\end{eqnarray*}

\begin{eqnarray*}
\sigma{}_{d}=\left(\begin{array}{cccccc}
	0 & 0 & 0 & 1 & 0 & 0\\
	0 & 0 & 1 & 0 & 0 & 0\\
	0 & 1 & 0 & 0 & 0 & 0\\
	1 & 0 & 0 & 0 & 0 & 0\\
	0 & 0 & 0 & 0 & 0 & 1\\
	0 & 0 & 0 & 0 & 1 & 0
\end{array}\right),  \quad \sigma{}_{d}C_3=\left(\begin{array}{cccccc}
	0 & 1 & 0 & 0 & 0 & 0\\
	1 & 0 & 0 & 0 & 0 & 0\\
	0 & 0 & 0 & 0 & 0 & 1\\
	0 & 0 & 0 & 0 & 1 & 0\\
	0 & 0 & 0 & 1 & 0 & 0\\
	0 & 0 & 1 & 0 & 0 & 0
\end{array}\right), \quad \sigma{}_{d}C^2_3=\left(\begin{array}{cccccc}
	0 & 0 & 0 & 0 & 0 & 1\\
	0 & 0 & 0 & 0 & 1 & 0\\
	0 & 0 & 0 & 1 & 0 & 0\\
	0 & 0 & 1 & 0 & 0 & 0\\
	0 & 1 & 0 & 0 & 0 & 0\\
	1 & 0 & 0 & 0 & 0 & 0
\end{array}\right),  \quad S_{6}^{2}=\left(\begin{array}{cccccc}
	0 & 0 & 0 & 0 & 0 & 1\\
	1 & 0 & 0 & 0 & 0 & 0\\
	0 & 1 & 0 & 0 & 0 & 0\\
	0 & 0 & 1 & 0 & 0 & 0\\
	0 & 0 & 0 & 1 & 0 & 0\\
	0 & 0 & 0 & 0 & 1 & 0
\end{array}\right)\\
\end{eqnarray*}

\begin{eqnarray*}
C_{2}^{\prime}=\left(\begin{array}{cccccc}
	1 & 0 & 0 & 0 & 0 & 0\\
	0 & 0 & 0 & 0 & 0 & 1\\
	0 & 0 & 0 & 0 & 1 & 0\\
	0 & 0 & 0 & 1 & 0 & 0\\
	0 & 0 & 1 & 0 & 0 & 0\\
	0 & 1 & 0 & 0 & 0 & 0
\end{array}\right),  \quad C_{2}^{\prime}C_3=\left(\begin{array}{cccccc}
	0 & 0 & 0 & 0 & 1 & 0\\
	0 & 0 & 0 & 1 & 0 & 0\\
	0 & 0 & 1 & 0 & 0 & 0\\
	0 & 1 & 0 & 0 & 0 & 0\\
	1 & 0 & 0 & 0 & 0 & 0\\
	0 & 0 & 0 & 0 & 0 & 1
\end{array}\right), \quad C_{2}^{\prime}C^2_3=\left(\begin{array}{cccccc}
	0 & 0 & 1 & 0 & 0 & 0\\
	0 & 1 & 0 & 0 & 0 & 0\\
	1 & 0 & 0 & 0 & 0 & 0\\
	0 & 0 & 0 & 0 & 0 & 1\\
	0 & 0 & 0 & 0 & 1 & 0\\
	0 & 0 & 0 & 1 & 0 & 0
\end{array}\right),  \quad E=\left(\begin{array}{cccccc}
	1 & 0 & 0 & 0 & 0 & 0\\
	0 & 1 & 0 & 0 & 0 & 0\\
	0 & 0 & 1 & 0 & 0 & 0\\
	0 & 0 & 0 & 1 & 0 & 0\\
	0 & 0 & 0 & 0 & 1 & 0\\
	0 & 0 & 0 & 0 & 0 & 1
\end{array}\right)
\end{eqnarray*}

The characters of $\Gamma$ in the Bloch basis are computed as:
\begin{align}
\chi^{\Gamma}(E) &=6, \quad \chi^{\Gamma}(I) = 0, \nonumber \\
\chi^{\Gamma}(C_{3}) &= \chi^{\Gamma}(C_{3}^{2}) = 2, \nonumber \\
\chi^{\Gamma}(S_{6}) &= \chi^{\Gamma}(S_{6}^{2}) = 0, \nonumber \\
\chi^{\Gamma}(C_{2}^{\prime}) &= \chi^{\Gamma}(C_{2}^{\prime}C_{3}) 
= \chi^{\Gamma}(C_{2}^{\prime}C_{3}^{2}) = 0, \nonumber \\
\chi^{\Gamma}(\sigma_{d}) &= \chi^{\Gamma}(\sigma_{d}C_{3}) 
= \chi^{\Gamma}(\sigma_{d}C_{3}^{2}) = 0. \nonumber
\end{align}

The decomposition of $\Gamma$ into its irreducible components can be obtained by  evaluating the multiplicities using Eq.~\eqref{Eq:A2}. We find that $\Gamma$ reduces to the direct sum
\begin{equation}
\Gamma  =A_{1g}\oplus E_{g}\oplus A_{1u}+E_{u}. \nonumber
\end{equation}

The projection operators $P_{\Gamma}$ for the irreducible representations (IRs) of $\Gamma$ were calculated using the matrix forms of the symmetry operations. The explicit forms are given below.

For the $A_{1g}$ representation:
\begin{equation}
\begin{aligned}
	P_{A_{1g}}= &\frac{1}{12} \Big[ \,
	\Gamma(E) + \Gamma(i) + \Gamma(C_{3}) + \Gamma(C_{3}^{2}) 
	+ \Gamma (C_{2}^{\prime}) \\
	& + \Gamma(C_{2}^{\prime}C_3) + \Gamma(C_{2}^{\prime}C^2_3) 
	+ \Gamma(S_{6}) + \Gamma(S_{6}^{2}) + \Gamma(\sigma_{d}) \\
	& + \Gamma(\sigma_{d}C_3) + \Gamma(\sigma_{d}C^2_3) \, \Big] \nonumber
\end{aligned}
\end{equation}

which reduces to
\begin{equation}
P_{A_{1g}}=\frac{1}{12}
\begin{pmatrix}
	2 & 2 & 2 & 2 & 2 & 2 \\
	2 & 2 & 2 & 2 & 2 & 2 \\
	2 & 2 & 2 & 2 & 2 & 2 \\
	2 & 2 & 2 & 2 & 2 & 2 \\
	2 & 2 & 2 & 2 & 2 & 2 \\
	2 & 2 & 2 & 2 & 2 & 2 \nonumber
\end{pmatrix}.
\end{equation}

The corresponding projected Bloch wave function is
\begin{eqnarray}
Z_{A_{1g}} &= & \frac{1}{6} \Big( 
e^{i\vec{k}\cdot\vec{\delta_{1}}} + e^{i\vec{k}\cdot\vec{\delta_{2}}} 
+ e^{i\vec{k}\cdot\vec{\delta_{3}}} + e^{i\vec{k}\cdot\vec{\delta_{4}}}  + e^{i\vec{k}\cdot\vec{\delta_{5}}} + e^{i\vec{k}\cdot\vec{\delta_{6}}} 
\Big). \nonumber\\
&=&2\left[\cos{k_x} + 2\cos{\left(\frac{k_x}{2}\right)}\cos{\left(\frac{\sqrt{3}k_y}{2}\right)}\right].
\label{Eq:ZA1g}
\end{eqnarray}

For the $A_{1u}$ representation:
\begin{equation}
P_{A_{1u}} = \frac{1}{12}
\begin{pmatrix}
	2 & -2 & 2 & -2 & 2 & -2 \\
	-2 & 2 & -2 & 2 & -2 & 2 \\
	2 & -2 & 2 & -2 & 2 & -2 \\
	-2 & 2 & -2 & 2 & -2 & 2 \\
	2 & -2 & 2 & -2 & 2 & -2 \\
	-2 & 2 & -2 & 2 & -2 & 2 \\ \nonumber
\end{pmatrix},
\end{equation}
with the projected wave function
\begin{eqnarray}
Z_{A_{1u}} &= &  \frac{1}{6}
\Big( e^{i\vec{k}\cdot\vec{\delta_{1}}} - e^{i\vec{k}\cdot\vec{\delta_{2}}} 
+ e^{i\vec{k}\cdot\vec{\delta_{3}}} - e^{i\vec{k}\cdot\vec{\delta_{4}}}  + e^{i\vec{k}\cdot\vec{\delta_{5}}} - e^{i\vec{k}\cdot\vec{\delta_{6}}} \Big). \nonumber\\
&=&2i\left[\sin{k_x} - 2\sin{\left(\frac{k_x}{2}\right)}\cos{\left(\frac{\sqrt{3}k_y}{2}\right)}\right].
\label{Eq:ZA1u}
\end{eqnarray}

For the two-dimensional $E_{g}$ representation:
\begin{equation}
P_{E_{g}} = \frac{2}{12}
\begin{pmatrix}
	2 & -1 & -1 & 2 & -1 & -1 \\
	-1 & 2 & -1 & -1 & 2 & -1 \\
	-1 & -1 & 2 & -1 & -1 & 2 \\
	2 & -1 & -1 & 2 & -1 & -1 \\
	-1 & 2 & -1 & -1 & 2 & -1 \\
	-1 & -1 & 2 & -1 & -1 & 2 \nonumber
\end{pmatrix},
\end{equation}
with the projected wave functions
\begin{eqnarray}
Z_{E_{g}^{(1/2)}} &=& \frac{2}{12}
\begin{pmatrix}
	2e^{i\vec{k}\cdot\vec{\delta_{1}}} - e^{i\vec{k}\cdot\vec{\delta_{2}}} - e^{i\vec{k}\cdot\vec{\delta_{3}}} 
	+ 2e^{i\vec{k}\cdot\vec{\delta_{4}}} - e^{i\vec{k}\cdot\vec{\delta_{5}}}-
	e^{i\vec{k}\cdot\vec{\delta_{6}}}\\
	-e^{i\vec{k}\cdot\vec{\delta_{1}}}+2e^{i\vec{k}\cdot\vec{\delta_{2}}} - e^{i\vec{k}\cdot\vec{\delta_{3}}} -e^{i\vec{k}\cdot\vec{\delta_{4}}} 
	+ 2e^{i\vec{k}\cdot\vec{\delta_{5}}} - e^{i\vec{k}\cdot\vec{\delta_{6}}}
\end{pmatrix}. \nonumber\\
&=&
\begin{pmatrix}
	2\cos{k_x} - 2\cos{\left(\frac{k_x}{2}\right)}\cos{\left(\frac{\sqrt{3}k_y}{2}\right)}\\
	- 2\sqrt{3}\sin{\left(\frac{k_x}{2}\right)}\sin{\left(\frac{\sqrt{3}k_y}{2}\right)}
\end{pmatrix}.
\label{Eq:ZEg}
\end{eqnarray}

For the two-dimensional $E_{u}$ representation:
\begin{equation}
P_{E_{u}} = \frac{2}{12}
\begin{pmatrix}
	2 & 1 & -1 & -2 & -1 & 1 \\
	1 & 2 & 1 & -1 & -2 & -1 \\
	-1 & 1 & 2 & 1 & -1 & -2 \\
	-2 & -1 & 1 & 2 & 1 & -1 \\
	-1 & -2 & -1 & 1 & 2 & 1 \\
	1 & -1 & -2 & -1 & 1 & 2 \nonumber
\end{pmatrix},
\end{equation}
with the corresponding projected wave functions
\begin{eqnarray}
Z_{E_{u}^{(1/2)}} &=& \frac{2}{12}
\begin{pmatrix}
	2e^{i\vec{k}\cdot\vec{\delta_{1}}} + e^{i\vec{k}\cdot\vec{\delta_{2}}} - e^{i\vec{k}\cdot\vec{\delta_{3}}} 
	- 2e^{i\vec{k}\cdot\vec{\delta_{4}}} - e^{i\vec{k}\cdot\vec{\delta_{5}}}+
	e^{i\vec{k}\cdot\vec{\delta_{6}}}\\
	e^{i\vec{k}\cdot\vec{\delta_{1}}}+2e^{i\vec{k}\cdot\vec{\delta_{2}}} +e^{i\vec{k}\cdot\vec{\delta_{3}}} -e^{i\vec{k}\cdot\vec{\delta_{4}}} 
	- 2e^{i\vec{k}\cdot\vec{\delta_{5}}} - e^{i\vec{k}\cdot\vec{\delta_{6}}}
\end{pmatrix}. \nonumber\\
&=&
\begin{pmatrix}
	2i\sin{k_x} + 2i\sin{\left(\frac{k_x}{2}\right)}\cos{\left(\frac{\sqrt{3}k_y}{2}\right)}\\
	2\sqrt{3}i\cos{\left(\frac{k_x}{2}\right)} \sin{\left(\frac{\sqrt{3}k_y}{2}\right)}
\end{pmatrix}.
\label{Eq:ZEu}
\end{eqnarray}

The Bloch wave spinors $Z_{\nu}$ deduced above are defined in the periodic lattice. We can expand them near the $\Gamma$ point while keeping upto $k^4$ order in both $k_x$ and $k_y$. We use the Taylor's expansion of $\sin(k_i)\approx k_i-\frac{k_i^3}{3!}+\mathcal{O}(k_i^5)$, and $\cos(k_i)\approx 1-\frac{k_i^2}{2}+\frac{k_i^4}{4!}-+\mathcal{O}(k_i^6)$. Then from Eqs.~\eqref{Eq:ZA1g}, \eqref{Eq:ZA1u}, \eqref{Eq:ZEg}, and \eqref{Eq:ZEu}, we obtain Eq.~2 in the main text, by collecting terms in terms of $k_{\pm}=k_x\pm ik_y$. Each term of the expansion can also be identified in terms of the orbital symmetry with angular momentum $(l,m_l)$, where the angular momentum of a term of the form $k_+^m k_-^n$ is defined as $l=m+n$ and $m_l=m-n$. Then we have 
\begin{eqnarray}
(l,m_l)&=&(1,\pm 1) : p_{x}={\rm Re}(k_+), \quad p_{y}={\rm Im}(k_+);\nonumber\\
&=&(2,0) :  s_{x^2+y^2} =k_+k_-=k_x^2 + k_y^2;\nonumber\\
&=&(2,\pm 2) : d_{x^2-y^2} = {\rm Re}(k_{+}^2)=k_x^2 - k_y^2, \quad  d_{xy}={\rm Im}(k_{+}^2)=k_xk_y;\nonumber\\
&=&(3,\pm 1) : Q_{x(x^2+y^2)}={\rm Re}(k_+^2k_-),\quad  Q_{y(x^2+y^2)}={\rm Im}(k_+^2k_-);\nonumber\\
&=&(3,\pm 3) :  f_{x(x^2-3y^2)}={\rm Re}(k_+^3)=k_x(k_x^2 - 3k_y^2),\quad  f_{y(y^2-3x^2)}={\rm Im}(k_+^3)=-k_y(k_y^2 - 3k_x^2); \nonumber\\
&=&(4,0) : K_{4x}={\rm Re}(k_{+}^4)= k_x^4+k_y^4-6k_x^2k_y^2,\quad   K_{4y}={\rm Im}(k_{+}^4)=4k_kk_y(k_x^2-k_y^2).
\end{eqnarray}

%
%The angular momentum of the term $k_+^m k_-^n$ is defined as $l=m+n$ and $m_l=m-n$. From this, we deduce the orbital symmetry of each term as $(l,m_l)=(1,\pm 1)$: $p_{x}={\rm Re}(k_+)$, $p_{y}={\rm Im}(k_+)$;  $(2,\pm 2)$: $d_{x^2-y^2} = {\rm Re}(k_{+}^2)=k_x^2 - k_y^2$  $d_{xy}={\rm Im}(k_{+}^2)=k_xk_y$; $(2,0)$: $s_{x^2+y^2} =k_+k_-=k_x^2 + k_y^2$; $(3,\pm 3)$: $f_{x(x^2-3y^2)}={\rm Re}(k_+^3)=k_x(k_x^2 - 3k_y^2)$,  $f_{y(y^2-3x^2)}={\rm Im}(k_+^3)=-k_y(k_y^2 - 3k_x^2)$;  $(3,\pm 1)$: $Q_{x(x^2+y^2)}={\rm Re}(k_+^2k_-)$, $Q_{y(x^2+y^2)}={\rm Im}(k_+^2k_-)$; $(4,0)$: $K_{4x}={\rm Re}(k_{+}^4)= k_x^4+k_y^4-6k_x^2k_y^2$,  $K_{4y}={\rm Im}(k_{+}^4)=4k_kk_y(k_x^2-k_y^2)$. 
%The coefficients $\alpha_i=\partial^{2i} Z_{A_{1g}}({\bf k})/\partial k_+^i\partial k_-^i\big|_{{\bf k}=0}$, and similarly for $\beta_i$, $\gamma_i$ and $\delta_i$}

\section{Spin Texture of the Other Kramers Pair}
\begin{figure}[h]
\centering
\includegraphics[height= 98 mm, width=185 mm,keepaspectratio]{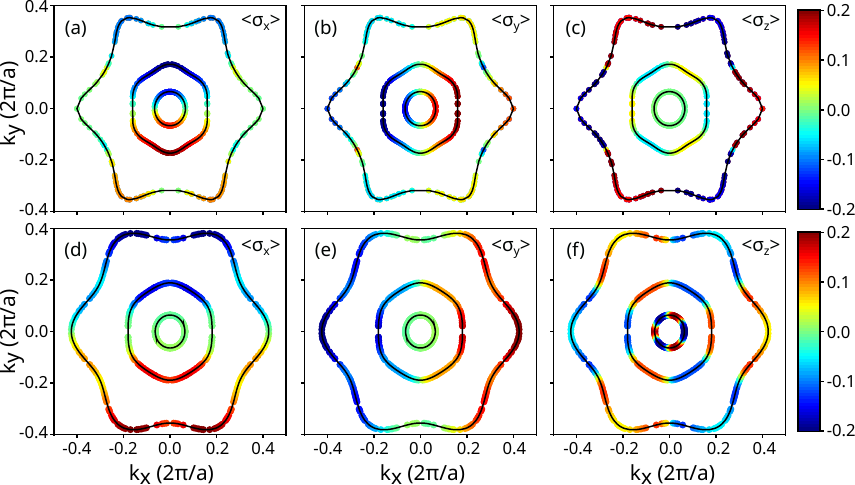}
\caption{ (a)–(c) Spin texture on the Fermi surface for the outer band, obtained from DFT calculations. The strength of the spin expectation values is indicated by the color. (d)–(f) Spin texture on the Fermi surface for the outer band, obtained from the $k \cdot p$ model Hamiltonian. The magnitude of the spin expectation values is similarly represented using a color.}
\label{SFig:4}
\end{figure}

In IrTe$_2$, the electronic bands are doubly degenerate due to the combined presence of inversion and time-reversal symmetries. Consequently, the spin expectation values associated with each degenerate band pair are equal in magnitude but opposite in direction. Figures~\ref{SFig:4}(a) and (c) show the spin expectation values obtained from first-principles DFT calculations, projected onto the Fermi surface. For comparison, Figs.~\ref{SFig:4}(d) and (f) present the corresponding spin expectation values calculated using the $\mathbf{k}\cdot\mathbf{p}$ model Hamiltonian and projected onto the Fermi surface for the other Kramers partner. The close agreement between the two approaches demonstrates the reliability of the effective model in capturing the essential spin textures of the system.
%%%%%%%%%%%%%%%%%%%%%%%%%%%%%%%%%%%%%%%%%%%%%%%%%%%%%%%%%%%%%%%%%%%%%%%%%%%%%%%%%%
%%%%%%%%%%%%%%%%%%%%%%%%%%%%%%%%%%%%%%%%%%%%%%%%%%%%%%%%%%%%%%%%%%%%%%%%%%%%%%%%%%

\section{Details of the pairing symmetry calculation}

We consider a multi-orbital Hubbard model given by 
\begin{equation}
H_{I}=%\sum_{{\bf R},\nu,\sigma}U_{\nu} n_{{\bf R},\nu,\sigma}n_{{\bf R},\nu,\bar{\sigma}}+
%\sum_{{\bf R},\nu,\nu',\sigma,\sigma'}V_{\nu,\nu'}n_{{\bf R},\nu,\sigma}n_{{\bf R},\nu',\sigma'} 
\sum_{{\bf q},m_l,m'_l,m_s,m_s'}V_{m_l,m'_l}n_{m_l,m_s}({\bf q})n_{m_l',m_s'}(-{\bf q}).
%+ J \sum_{i \neq j}\textbf{S}_i\cdot \textbf{S}_j+J' \sum_{i \neq j,\sigma}c^\dagger_{i\sigma}c^\dagger_{i \bar{\sigma}}c_{j\bar{\sigma}}c_{j \sigma}
\label{eq-Hubbard}
\end{equation}
Here, $V_{m_l,m'_l}$ incorporates both intra-orbital ($m_l=m'_l$, $m_s\ne m'_s$) contribution, which we denote by $U_{m_l}$, and the inter-orbital contribution $m_l\ne m'_l$ , where $m_s$ and $m_s'$ can be the same spin or different. $n_{m_l,m_s}({\bf q})=\sum_{\bf k}c^{\dagger}_{m_l,m_s}({\bf k}+{\bf q})c_{m_l,m_s}({\bf k})$ is the density matrix between different orbitals $(m_l,m_l')$, spins $(m_s,m_s')$, and a momentum transfer of ${\bf q}$. The operators $c^{\dagger}_{m_l,m_s}({\bf k})$ correspond to the creation of electrons in the $m_l,m_s$ basis with momentum ${\bf k}$. All momentum summations are assumed to be weighted with the Brillouin zone volume. We ignore the spin exchange and pair exchange interactions as they possess less contribution. 

Now, performing the standard RPA-level perturbation expansion of $H_I$,\cite{Scalapino1986, Schrieffer1989, Sigrist1991, Takimoto2004, Mazin2008,Graser2009, Scalapino2012, Das2014}, and collecting the terms which give an effective pairing interaction $V_{\nu_l}^{\nu_s}({\bf q})$, where $\nu_l$ corresponds to the three irreps of the orbitals $(A_{1g}$, $A_{2g}$, and $E_g$) that are defined in {Eq.~5 } in the main text, and $\nu_s=\pm$ corresponds to the spin-triplet and spin-singlet irreps. Here, we are not using the irreps of the Bloch phase $Z({\bf q})$, although it can be easily done. Then the effective interaction term with the RPA correction can be written in terms of the pairing fields $\Psi({\bf k})$,  in {(Eq.~4)},  in the $\nu_l,\nu_s$ channels as
\begin{equation}
H_{I,{\rm RPA}} = \sum_{\nu_l,\nu_s}\sum_{{\bf k},{\bf q}} V_{\nu_l,\nu'_l}^{\nu_s,\nu'_s} (\textbf{q})\Psi^{\dagger}_{\nu_l,\nu_s}(\textbf{k})\Psi_{\nu'_l,\nu'_s}(\textbf{k}+\textbf{q}).
%= \sum_{\nu,\nu',\sigma,\sigma'}\sum_{{\bf k},{\bf q}} \Gamma_{\nu\nu'}^{\sigma\sigma'} (\textbf{q})c^{\dagger}_{\nu,\sigma}(\textbf{k})c^{\dagger}_{\nu',\sigma'}(-\textbf{k}) c_{\nu',\sigma'}(-\textbf{k}-\textbf{q}) c_{\nu,\sigma}(\textbf{k}+{\bf q}).
\label{Eq:RPAPair}
\end{equation}
In the second quantization, the pair fields are written as  $$\Psi_{\nu_l,\nu_s}(\textbf{k})=\mathsf{P}_{-}\left[\sum_{m_l,m_l'}\sum_{m_s,m_s'}U_{\nu_l}^{m_l,m_l'}U_{\nu_s}^{m_s,m_s'}\left(c_{m_l,m_s}(-\textbf{k}) c_{m_l',m_s'}(\textbf{k})\right)\right]$$. Here $U_{\nu_l}^{m_l,m'_l}$ gives the transformation from the two-orbital product state to the irreps state, and   $U_{\nu_s}^{m_s,m_s'}$ gives the same for the spins. $\mathsf{P}_-$ ensures the pair fields remain antisymmetric under the change of orbitals, spins, and also momentum. 

The pair interaction $V_{\nu_l,\nu'_l}^{\nu_s,\nu'_s}$ in this form is assumed to mix different irreps in both spin and orbital channels, in general. Such mixing can occur when the interaction term allows spin or pair exchange terms. Also, because we have not used the irreps of the Bloch phase above, at a given momentum, different orbital irreps can, in principle, mix. In the actual calculation, we, however, find that the different irreps do not mix. The pair interaction term can be  transformed to the orbital basis in a similar way as $V_{\nu_l,\nu'_l}^{\nu_s,\nu'_s}(\textbf{q})=\sum_{\{m_l\}}U_{\nu_l}^{m_l,m'_l}U_{\nu'_l}^{m''_l,m'''_l}\left(V_{\{m_l\}}^{\nu_s,\nu'_s}(\textbf{q})\right)$. $\{m_l\}=(m_l,m'_l,m''_l,m'''_l)$. We only have $\nu_s=\nu_s'=\pm$ terms for the spin singlet and spin triplets, and no singlet-triplet mixing is found here. Therefore, we have  the pairing potential in the orbital basis is then written as 
\begin{eqnarray}
V^{\nu_s}_{\{m_l\}}(\textbf{q}) &=& - \dfrac{\nu_s}{2}\Big[\eta_{\nu_s}U_s \chi_s(\textbf{q})U_s  +\nu_s U_c\chi_c(\textbf{q})U_c -\nu_s (U_s + U_c)\Big]_{\{m_l\}}.
%\Gamma_{}(\textbf{q}) &= \dfrac{1}{2}\left[3\tilde{U}_s\tilde{\chi}_s(\textbf{q})\tilde{U}_s - \tilde{U}_c\tilde{\chi}_c(\textbf{q})\tilde{U}_c+\tilde{U}_s+\tilde{U}_c)\right] \\
%\tilde{\Gamma}_{\uparrow\uparrow}(\textbf{q}) &= -\dfrac{1}{2}\left[\tilde{U}_s\tilde{\chi}_s(\textbf{q})\tilde{U}_s + \tilde{U}_c\tilde{\chi}_c(\textbf{q})\tilde{U}_c-\tilde{U}_s-\tilde{U}_c)\right]
\end{eqnarray}
where $\eta_{\pm}=3$ for $\nu_s=-1$ and 1 for $\nu_s=+1$. The subscripts $s$ and $c$ refer to the spin and charge density fluctuation channels, respectively, and the corresponding on-site interaction matrices are represented by $U_{s/c}$. In the present case $U_s=U_c$ and this matrix has intra-orbital interaction $U_{m_l}$ in the diagonal entries and inter-orbital terms $V_{m_l,m'_l}$ in the off-diagonal entries. 
$\chi_{s/c}$ are the RPA density-density correlators for the spin and charge density channels. The RPA susceptibilites are related to the Lindhard susceptibility $\chi_0$ as:
\begin{equation}
\chi_{s/c}(\textbf{q})=\chi_{0}(\textbf{q})\left({\mathbb{I}}\mp{U}_{s/c}{\chi}_0(\textbf{q})  \right)^{-1}
\label{eq-chi_rpa}
\end{equation}
where ${\mathbb{I}}$ is the unit matrix.  %The strong FS nesting features captured within the Lindhard susceptibility in Eq. (\ref{eq-chi0}) are automatically translated to strong peaks in the RPA susceptibilities in Eq. (\ref{eq-chi_rpa}). The RPA denominator of the spin susceptibility, having value $<1$, enhances the FS nesting strength in the bare susceptibility $\tilde{\chi}_0(\textbf{q})$. On the other hand, the RPA denominator for the charge channel is $>1$ suppressing the charge fluctuations. 

\subsubsection{Superconducting pairing symmetry}
%Eq \ref{eq-Hubbard} gives the pairing interaction for pairing between orbitals. However, we solve the BCS gap equation in the band basis. To make this transformation, we make use of the unitary transformation $c_{ij}\rightarrow \sum_{\nu}\mathcal{U}^i_{\nu} l\nu\sigma$ for for all $\textbf{k}$ and spin $\sigma$.  With this substitution we obtain the pairing interaction Hamiltonian in the band basis as
%\begin{equation}
%H_{int} \approx \dfrac{1}{\Omega^2_{BZ}}  \sum_{\nu,\nu'}\sum_{k,q,\sigma,\sigma'} \Gamma'_{\nu,\nu'} (\textbf{k},\textbf{q})  \times l^{\dagger}_{\nu,\sigma}(\textbf{k})l^{\dagger}_{\nu,\sigma'}(-\textbf{k}) l_{\nu',\sigma'}(-\textbf{k}-\textbf{q}) l_{\nu',\sigma}(\textbf{k}+\textbf{q})
%\end{equation}
%The same equation holds for both singlet and triplet pairing and thus henceforth we drop the corresponding symbol for simplicity. The band pairing interaction $\Gamma'_{\nu\nu'}$ is related to the corresponding orbital one as 
%\begin{equation}
%\Gamma'_{\nu\nu'}(\textbf{k},\textbf{q})=\sum_{i,j,l,m}\Gamma_{i,j}^{l,m}(\textbf{q})\phi_i^{\nu\dagger}(\textbf{k})\phi_j^{\nu\dagger}(-\textbf{k})\phi_l^{\nu'}(-\textbf{k}-\textbf{q})\phi_m^{\nu'}(\textbf{k}+\textbf{q})
%\end{equation}
From Eq.~\eqref{Eq:RPAPair}, we define the SC gap as
\begin{equation}
\Delta_{\nu}(\textbf{k})= -\sum_{\nu',\textbf{q}}V_{\nu,\nu'}(\textbf{q})\left\langle \Psi_{\nu'}(\textbf{k}+\textbf{q})\right\rangle,
\label{eq-del_nu-1}
\end{equation}
where the expectation value is taken over the BCS ground state. In the above we used a composite symbol $\nu=(\nu_l,\nu_s)$. In the limit, $T\rightarrow0$ we have $\left\langle \Psi_{\nu}(\textbf{k}+\textbf{q})\right\rangle \rightarrow \lambda^{-1}\Delta_{\nu}(\textbf{k}+{\bf q})$ with $\lambda$ as the SC coupling constant. Substituting this in Eq. (\ref{eq-del_nu-1}), we get {Eq.~6} in the main text.

\end{document}